\documentclass{article}

\usepackage{PRIMEarxiv}

\usepackage[utf8]{inputenc} 
\usepackage[T1]{fontenc}    
\usepackage[hidelinks]{hyperref}       
\usepackage{url}            
\usepackage{booktabs}       
\usepackage{amsfonts}       
\usepackage{nicefrac}       
\usepackage{microtype}      
\usepackage{lipsum}
\usepackage{fancyhdr}       
\usepackage{graphicx}       
\graphicspath{{media/}}     
\usepackage{subcaption}

\usepackage{algorithm}
\usepackage{algpseudocode}
\usepackage[english]{babel}

\title{Kolam Simulation using Angles at Lattice Points
}

\author
  {Tulasi Bharathi, Shailaja D Sharma, Nithin Nagaraj \\
  National Institute of Advanced Studies \\
  Indian Institute of Science Campus\\
  Bangalore 560 012, India.\\
  \texttt{\{bharathitulasi3@gmail.com, shailaja@nias.res.in, nithin@nias.res.in\}}
}

\begin{document}
\maketitle

\begin{abstract}
Kolam is a ritual art form practised by people in South India and consists of rule-bound geometric patterns of dots and lines. Single loop Kolams are mathematical closed loop patterns drawn over a grid of dots and conforming to certain heuristics. In this work, we propose a novel encoding scheme where we map the angular movements of Kolam at lattice points into sequences containing $4$ distinct symbols. This is then used to  simulate single loop Kolam procedure via turtle moves in accordance with the desired angular direction at specific points. We thus obtain sequential codes for Kolams, unique up to cyclic permutations. We specify the requirements for the algorithm and indicate the general methodology. We demonstrate a sample of Kolams using our algorithm with a software implementation in Python.
\end{abstract}


\section{Introduction}
Kolam is a collection of classes of predominantly geometric designs widely prevalent in South India, drawn on the freshly washed threshold of homes at dawn, using dry rice powder or soapstone powder or using a paste of rice or clay in water. Women practitioners draw Kolams from memory and practice, without reference to a sample image. It is well-known that Kolam drawing requires focus and practice as the possibilities of error are overwhelming. The Kolams we discuss here are drawn with reference to a grid of dots. In drawing single-loop Kolams, the artist has to make the correct decision about the direction of the loop in the vicinity of each point in the grid, failing which the Kolam cannot be completed and must be fully erased and re-drawn. Competent Kolam artists never make a mistake in executing a Kolam pattern, as the floor is cleaned before the Kolam is drawn upon it and erasure of the Kolam at any point damages its appearance.
Kolams are thus `finished designs' in the sense that they are not improvised, although improvisation is not ruled out. Artists must be familiar with the Kolam they propose to execute. Thus, with each Kolam can be identified a unique sequence of moves, assuming a given starting point. Kolam artists may start the Kolam at any point they desire, but they have the entire image in their minds and execute the moves unerringly so as to obtain the desired final image. In this paper, we propose a deterministic algorithm which simulates this procedure. For this purpose, we establish a starting point for the Kolam to be drawn and translate the Kolam to a sequence of angular movements connected by smooth lines. The resulting sequence uniquely, up to cyclic permutations, describes the given Kolam. 
\section{Literature Survey}
Kolam construction received the attention of scholars from computer science and computer graphics and an early review of extant methods was provided by \cite{Ascher2002}. An updated review of computational methods of Kolam generation was undertaken by \cite{Akhilesh}. A significant number of the methods for Kolam construction deconstruct the final image and propose algorithms to generate the Kolam by concatenating the constituent units together correctly \cite{SIROMONEY197463}. Finite, repetitive and recursive Kolam structures, based on known traditional Kolams, are discussed by \cite{SIROMONEY197463}, based on array rewriting methods discussed in \cite{SIROMONEY1973447}. It is relevant to note that of the large number of Kolam-like patterns that can be drawn, very few would however qualify as Kolams. There is a notion of ``correctnes'' or satisfaction of a set of heuristics, pertaining to symmetry and unicursality, which restricts the possibilities for a Kolam. \cite{Yanagisawa2007} computes the number of isomorphic Kolams for particular dot grid dimensions. Identification of the typical repetitive patterns in the Kolam is a key part of the algorithmic procedures proposed by \cite{SIROMONEY197463}. Recursive algorithms are used to generate sequences of Kolams of the same pattern. The variety and scope of such approaches can also be gauged from the discussion in \cite{Nagata2023}. 
\par Lindenmayer language and turtle moves have been used by various authors who recognized the nested patterns in Kolam \cite{LINDENMAYER1968280}. Lindenmayer language is suitable for generative structures and was originally proposed for modeling the development of plants. The addition of turtle moves and turtle graphics enabled the depiction of both plant-like and also closed loop structures. Angle of rotation for such Kolams is a fixed parameter, which may take positive or negative directions. It is noteworthy that while such algorithms do generate the desired final Kolam images, they do not necessarily reproduce the human act of Kolam. In other words the sequence of lines constructed may differ although the final image is the same.
\par 
The methods discussed by \cite{Yanagisawa2007} refer to the procedural nature of Kolam production and propose algorithms for Kolam generation by coding the unit tile enclosing each Kolam dot. The method is based on classifying the finite number of patterns that are generated around a Kolam dot by the looping Kolam line. It may be said that the sequential patterns so created are based entirely on local considerations, i.e. the desired pattern around a given Kolam dot. The digital sequence so generated is unique up to cyclic permutations and the starting point is selected arbitrarily or by setting up a convention, in a fixed orientation. The present algorithm shares these features. \cite{Yanagisawa2007} also discusses N-lines, which reflect the overall (global) structure of the nodes and tree branches in Kolams. 
\par 
We observe that the human act of Kolam drawing is based on decisions about direction of movement, taken at various points (lattice points) as the looping curve traverses the Kolam dot grid, The present paper therefore proposes an algorithm that mimics the human decision-making behaviour in drawing Kolam. The decision-making takes into account the global requirements of the Kolam as well as local requirements, simultaneously. By `global' we mean here the overall shape of the Kolam and by `local' just the shape of the Kolam in the vicinity of a given Kolam dot. Our algorithm anticipates the overall structure of the curve in making a local determination. 
\par
We first discuss the methodology involved in Kolam simulation using angles at lattice points. Secondly, the data pre-processing required for simulation is given. Thirdly, the results of Kolam simulation are presented.

\section{Terminology} \label{Terminology}
 In order to discuss the algorithmic procedure, we introduce some terminology. We may borrow terminology from graph theory, lattices or knot theory in mathematics, as well as others. By \textit{Kolam}, we refer to the a stylized geometrical pattern of rigid or flexible lines, which is a distinctive cultural artefact of Southern India (although it may be practised elsewhere as well). A \textit{Kolam dot grid} is the mathematical arrangement of dots, which is mandatory for drawing certain classes of Kolams, which we shall refer to as \textit{dot-Kolams}. We shall refer to the dots of this underlying grid as \textit{Kolam dots}. The Kolams discussed in this paper are restricted to looping Kolams, or Line-Around-Dots (LAD) Kolams \cite{Nagata} in which the solid lines of the Kolam curve around the Kolam dots but never go through a Kolam dot. Traditionally these are called \textit{chuzhi} Kolam, the Tamil technical word referring to a loop.  
\par We have a reference plane on which the dot grid is laid out and the perpendicular axes of this plane may be identified with the cardinal directions. The square dot grid, aligned with the cardinal directions, will be referred to as a $n\times n$ dot grid. In the case that the square dot grid is inclined at $45$ degrees to the cardinal directions, we treat it as a rotated square Kolam. Rhombic Kolam \cite{Yanagisawa2007} has a dot grid with a different arrangement of dots, with successive rows containing successive odd numbers of dots. Such a dot grid will be referred to using the notation $1-n-1$, where $n$ refers to the maximal width of the Kolam and is usually an odd number. Both square and rhombic Kolams are treated in this paper (see Fig. \ref{F5}) . 
\par A single-loop Kolam is a Kolam that is in the form of a closed loop that is entwined with the Kolam dot grid in LAD mode. Reference to the importance of this type of Kolam is also found in the earliest Western compilation of Kolams written by John Layard \cite{John}.  The traditional technical term for such closed loop Kolams is \textit{Brahmamudi}, which is suggestive of the importance attached to single closed loops for which the starting/end point is ambiguous. The term \textit{mudi} refers to a top-knot, and the connection with the mathematical concept of knots can also be observed. Here, a knot is understood to be a $2$-dimensional tie whose ends cannot be identified. For a set of rules on the allowed and disallowed patterns in a looping Kolam, the reader is referred to \cite{Yanagisawa2007}.
\par Given a Kolam dot grid, we impose upon it a lattice structure such that every Kolam dot is enclosed within a unit cell, with the lattice points lying in the cardinal directions around it (see Figs. \ref{F3}, \ref{F4}, \ref{F5}). We refer to the lattice point where the Kolam begins to be drawn as \textit{starting point}. We characterize the Kolam by the sequence of angles made at the lattice points with respect to the axes of the plane, assuming a certain starting point. The Kolam here goes through the lattice points (but not through the Kolam dots). The solid line forms smooth turns or goes straight through at each lattice point (see Figs. \ref{F1}, \ref{F2}). Thus, the Kolam is represented as a sequence of angles, once information about the dot grid ($n\times n$  or $1-n-1$) and starting point are given. A convention for starting point is adopted; for square Kolams it shall be the $270$° lattice point (lowest point) of the unit cell of the top-most and left-most Kolam dot; for rhombic Kolams it shall be the  $270$° lattice point of the unit cell of the left-most Kolam dot (see Fig. \ref{F8}). These roughly correspond to  some natural starting points for right-handed Kolam artistes, but the choice is otherwise arbitrary.
\par The lattice point at which the Kolam is currently being drawn is referred to as the \textit{current point}. The three succeeding lattice points through which the Kolam must pass after the current point are referred to as the \textit{first, second} and \textit{third} points respectively and the numbering is with respect to the current point. 

\section{Methodology}
\label{sec:headings}

\begin{figure}
\centering
    \includegraphics[scale=0.5]{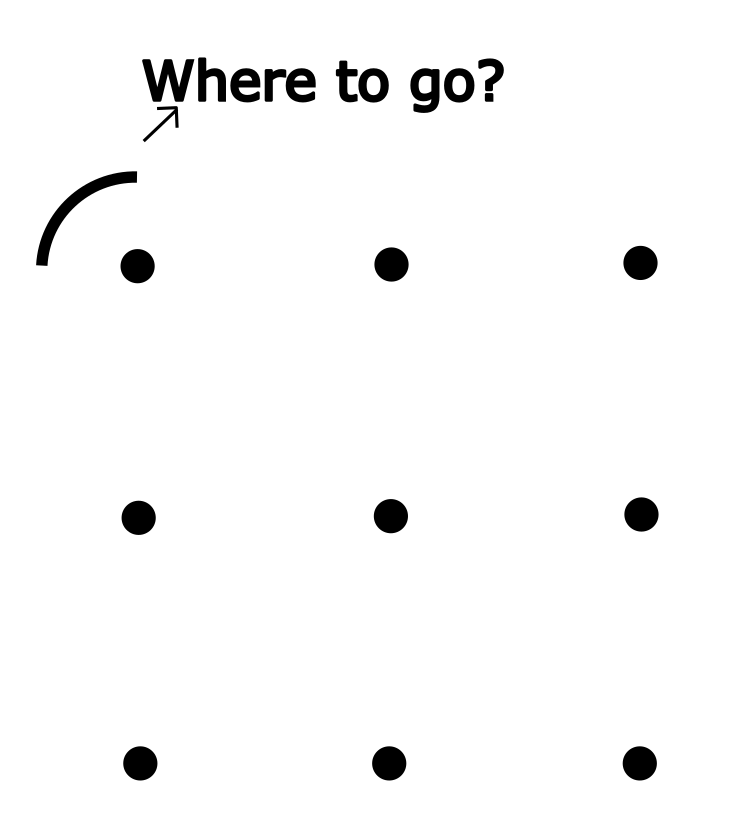}
    \caption{Deciding where to go while drawing a Kolam - Part 1.}
\label{F1}
\end{figure}

\begin{figure}
\centering
    \includegraphics[scale=0.5]{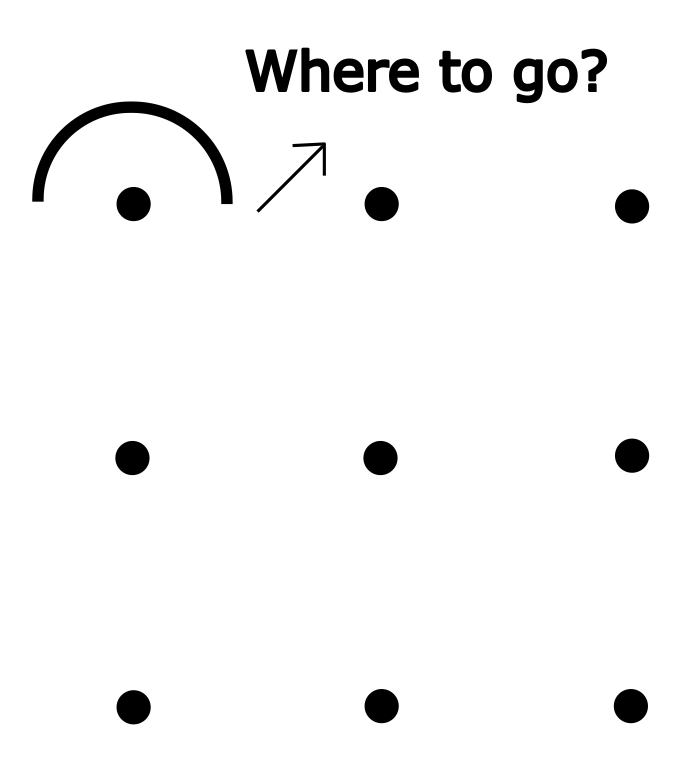}
    \caption{Deciding where to go while drawing Kolam - Part 2.}
\label{F2}
\end{figure}

 Kolam selection is made before commencing the act of drawing. Kolam drawing is initiated by specifying the corresponding dot grid. (Note that several different Kolams may have the same dot grid.) For the purposes of the algorithm, an abstract lattice structure is constructed over the Kolam dot grid by placing a rotated square unit cell around each Kolam dot. The Kolam is drawn starting at a lattice point. Thereafter, a decision as to the direction in which to extend the line has to be taken. Mathematically speaking, there are infinite possibilities, but Kolam construction entails discretizing these infinite decisions into just 4 possible outcomes, viz. $45$°, $135$°, $-45$°, $-135$°. Note that the angular values are established with reference to the orthogonal axis system to which the Kolam dot grid is aligned. Figures \ref{F1} and \ref{F2} depict the idea of decision making while drawing a Kolam. The shape of the Kolam line between two lattice points may be convex or concave, as can be seen from the diagram. This choice of shape requires anticipating two steps at each decision-making point, thus the algorithm will look-ahead to maximum three subsequent Kolam points at each stage of its development. \par


\begin{figure}
\centering
    \includegraphics[scale = 0.9]{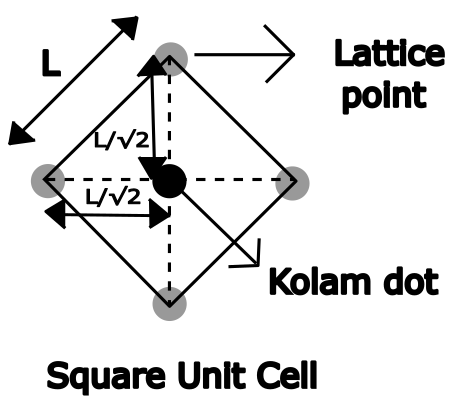}
    \caption{A square unit cell imposed on a Kolam dot.}
\label{F3}
\end{figure}
\begin{figure}
\centering
    \includegraphics[scale = 1]{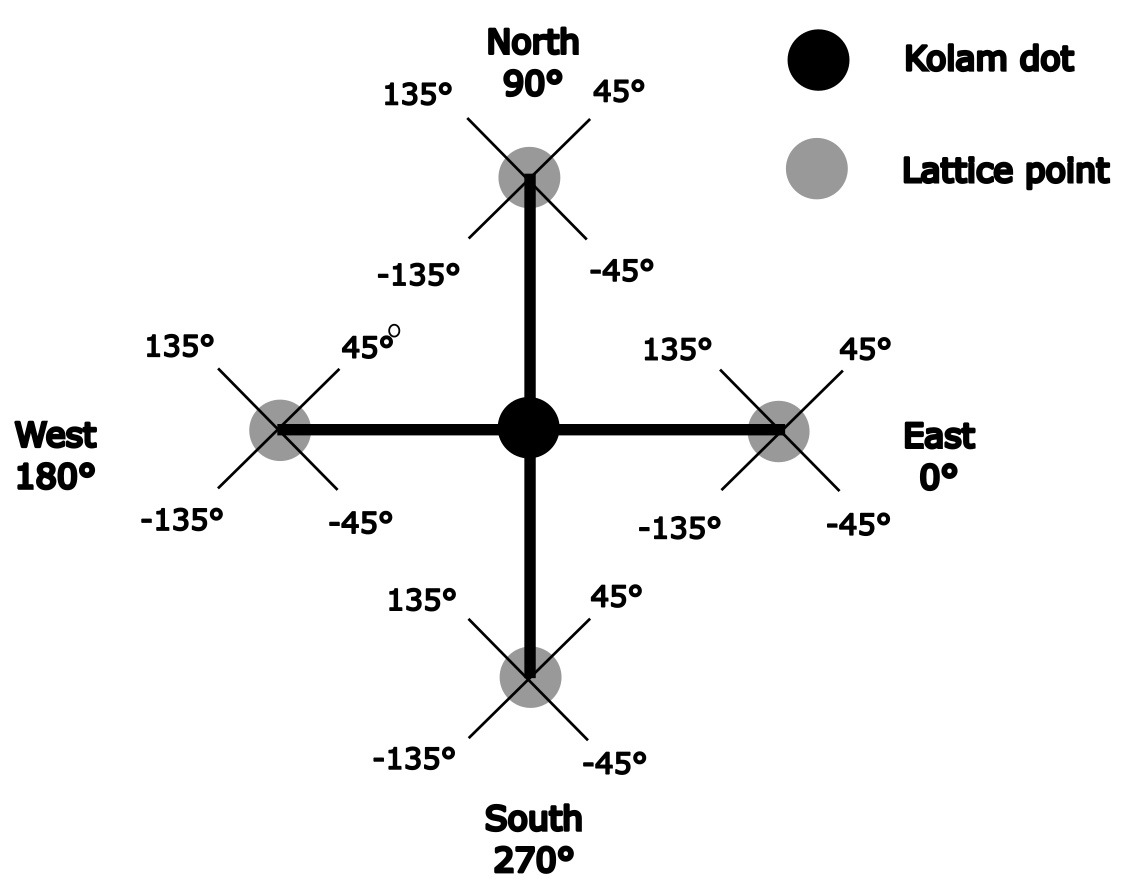}
    \caption{Cardinal directions of a Kolam dot and intercardinal directions of a lattice point.}
\label{F4}
\end{figure}
\begin{figure}
\centering
    \includegraphics[scale = 0.9]{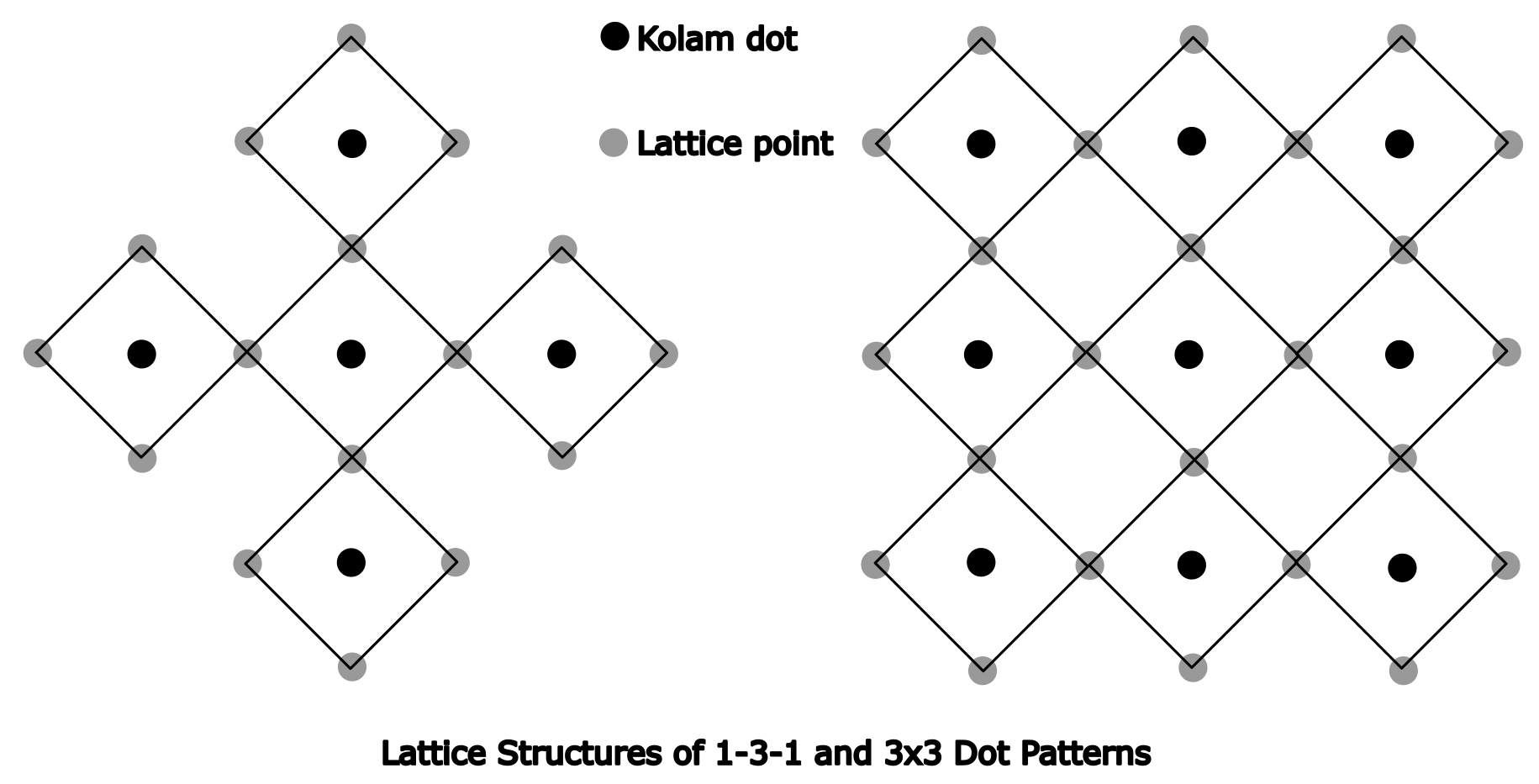}
    \caption{Lattice structures imposed on $1-3-1$ and $3\times3$ dot patterns.}
\label{F5}
\end{figure}

\begin{figure}
\centering

    \includegraphics[scale=0.9]{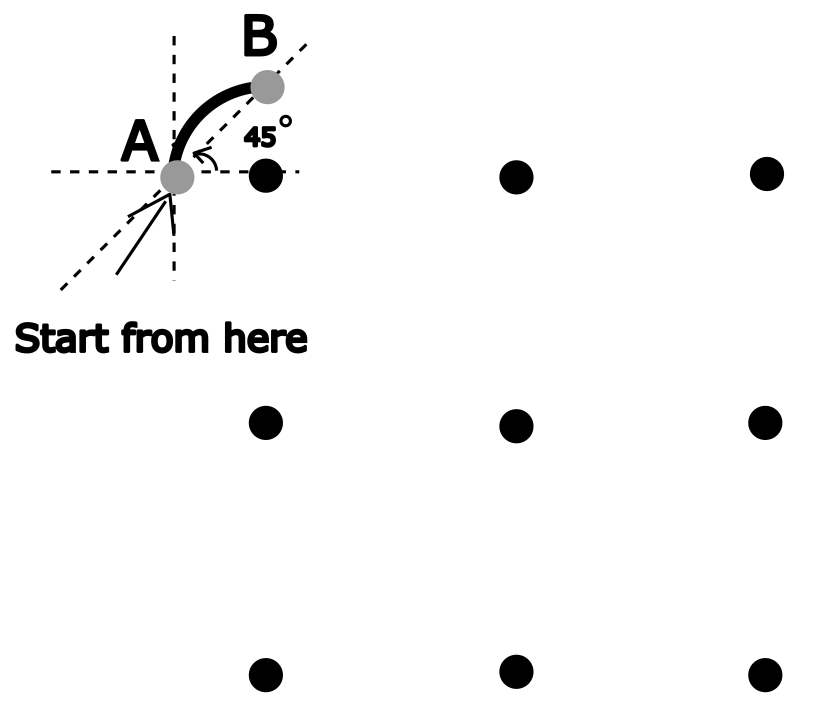}
    \caption{Direction of point $B$ from point $A$.}
\label{F6}    
\end{figure}

\begin{figure}
\centering
    \includegraphics[scale=0.8]{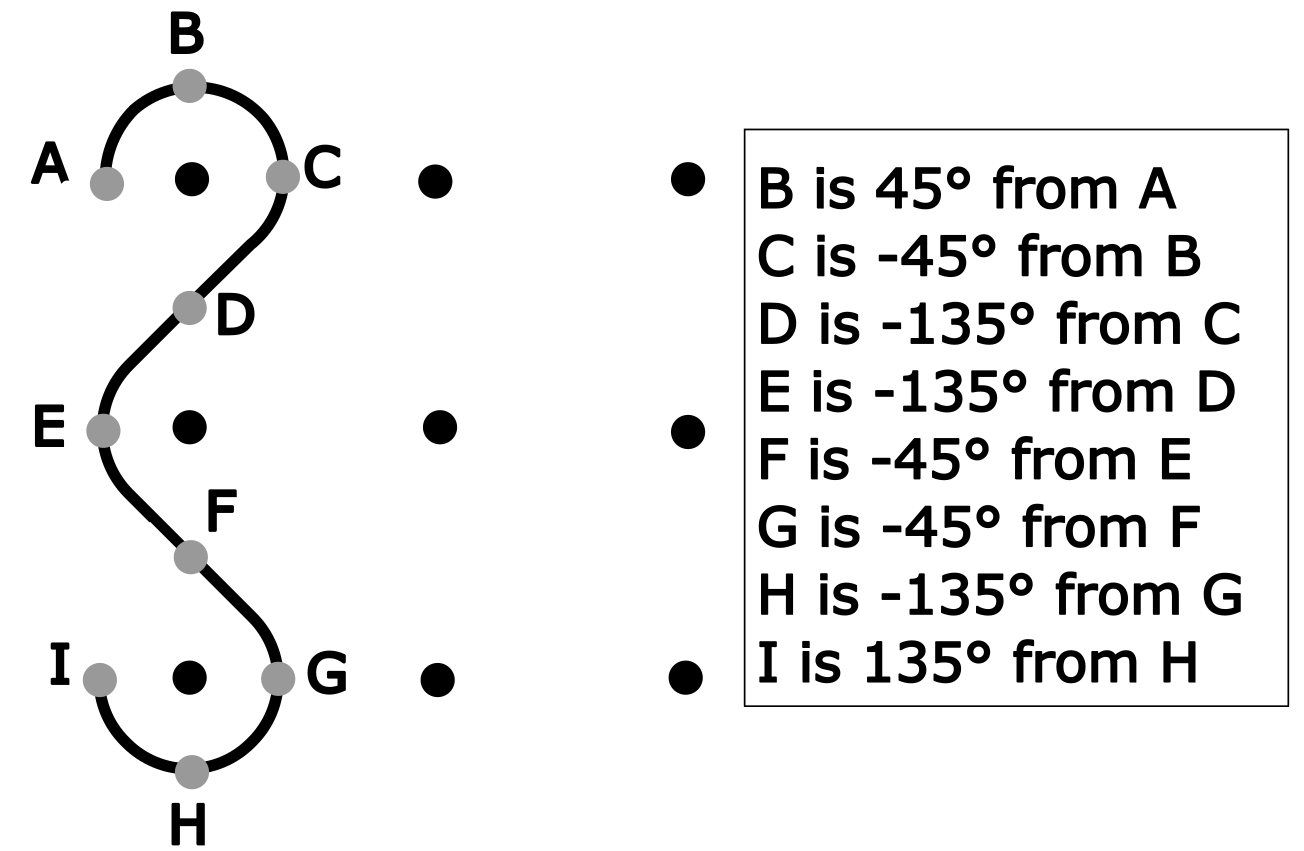}
    \caption{Direction of a point with respect to its predecessor point.}

\label{F7}
\end{figure}

\begin{figure}
\centering
    \includegraphics[scale = 0.9]{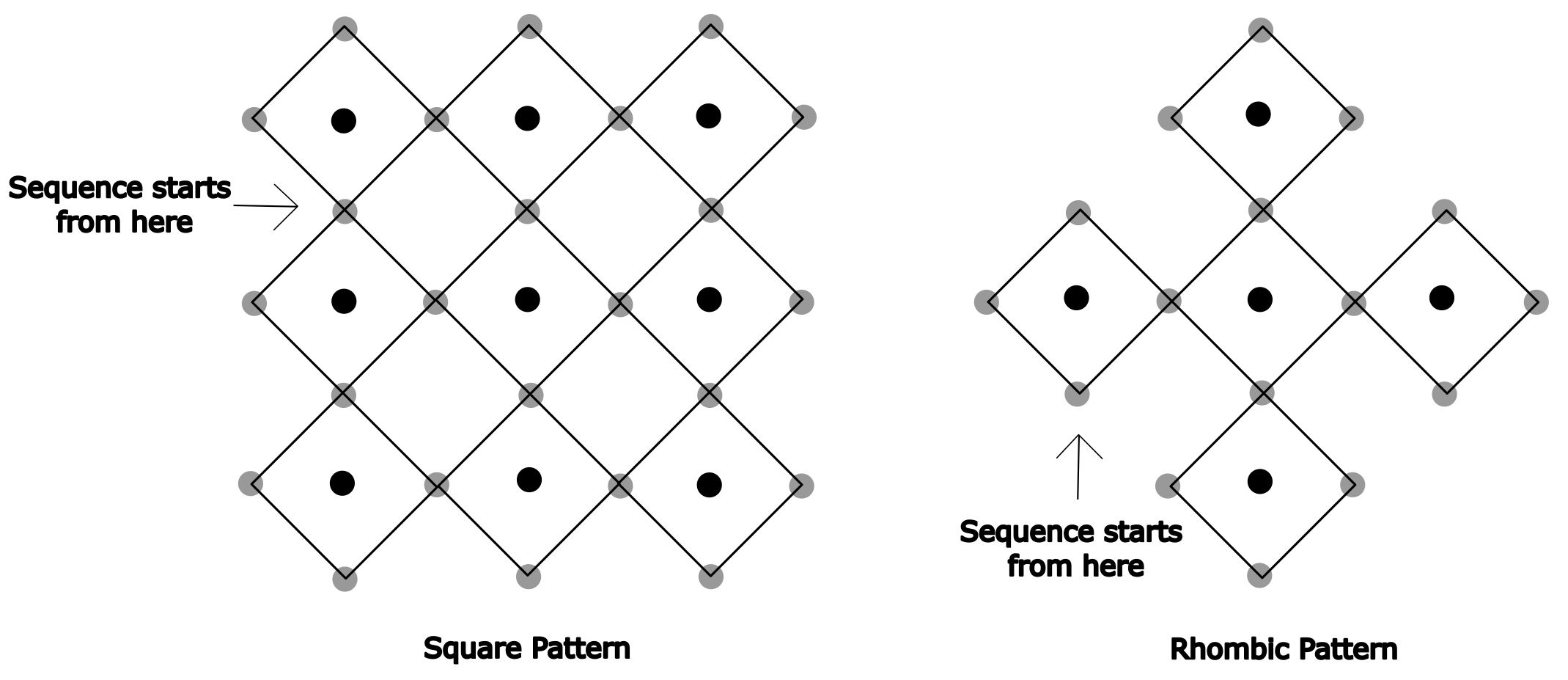}
    \caption{Convention for sequence start-point of square and rhombic Kolams.}
\label{F8}
\end{figure}

 Suppose that the Kolam dot grid has points separated by a distance $L\sqrt{2}$. Then the unit cell in the lattice has sides of length $L$. Starting at a particular lattice point, for every distance of length $L$ along the edges of the lattice cells we make a selection one out of the four angular directions given above. The angle information at the lattice points of a Kolam which determines the sequential path of a Kolam is not arbitrary and is pre-determined from the shape of the Kolam to be depicted. In other words, the Kolam is fully described by the sequence of angles, up to isomorphisms.  The choice of angle at lattice points and how to draw a Kolam using it are shown in Figures \ref{F3} - \ref{F8}. \par

\par

\section{Data Pre-processing} \label{Data Pre-processing}
\subsection{Kolam encoding} \label{Kolam encoding}
\begin{figure}
\centering
    \includegraphics[scale = 1.8]{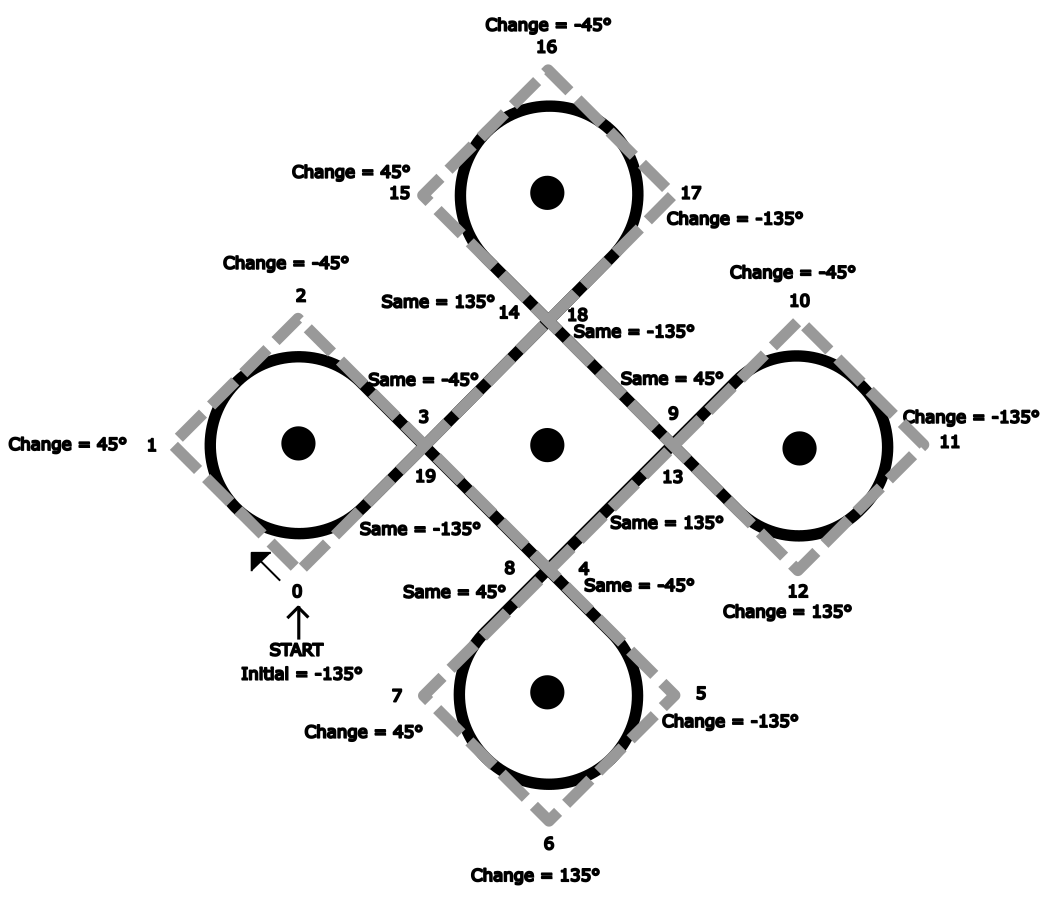}
    \caption{A $1-3-1$ Kolam with angle changes while drawing.}
\label{F9}
\end{figure}

\begin{figure}
\centering
    \includegraphics[scale=2]{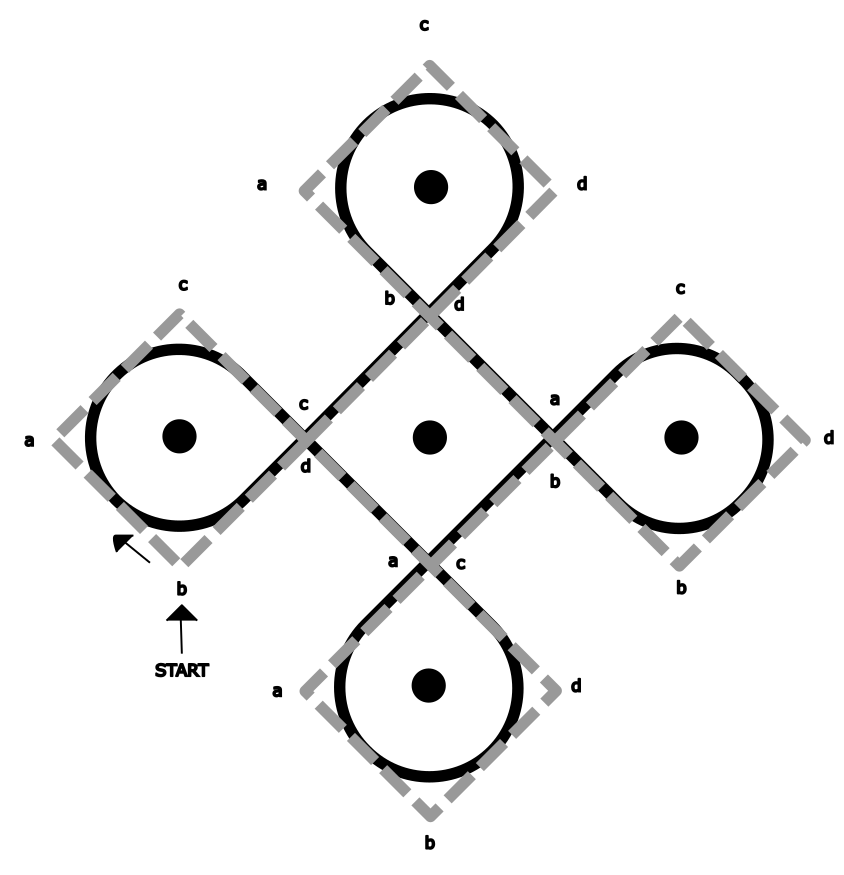}
    \caption{Angle changes in Fig.~\ref{F9} replaced with their corresponding symbols.}
\label{F10}
\end{figure}

Every lattice point is like an origin in an $XY$-plane, in the sense that it has no pre-defined direction. At each lattice point, we can assume four inter-cardinal directions as possibilities, as shown in Figure \ref{F4}. We represent the 4 angles, $45$°, $135$°, $-45$°and $-135$°(i.e. the inter-cardinal directions) by the symbols $a$, $b$, $c$ and $d$ respectively, for ease of use as well as generalization purposes (see Table \ref{T1}). Now, we claim that every dot Kolam of the \textit{Brahmamudi} (single closed loop) variety can be represented using a finite sequence of elements from this alphabet (although the converse is not true). Further, each such sequence uniquely describes the corresponding Kolam, up to isomorphisms. A $1-3-1$ Kolam with its angle information in degrees is shown in Figure \ref{F9}.  The angles of $1-3-1$ Kolam are replaced with their corresponding symbols and is shown in Figure \ref{F10}. The Kolams and their corresponding sequences are given in Table \ref{T2} and \ref{T3}.\par

 \begin{table}
 \centering
\caption{Coding lattice angles}
\vspace{0.1in}
\begin{tabular}{cc}
\hline
Angle & Symbol \\ \hline
$45$°   & $a$      \\
$135$°  & $b$      \\
$-45$°  & $c$      \\
$-135$° & $d$      \\ \hline
\end{tabular}
\label{T1}
\end{table}

\begin{figure}
\centering
    \includegraphics[scale=1]{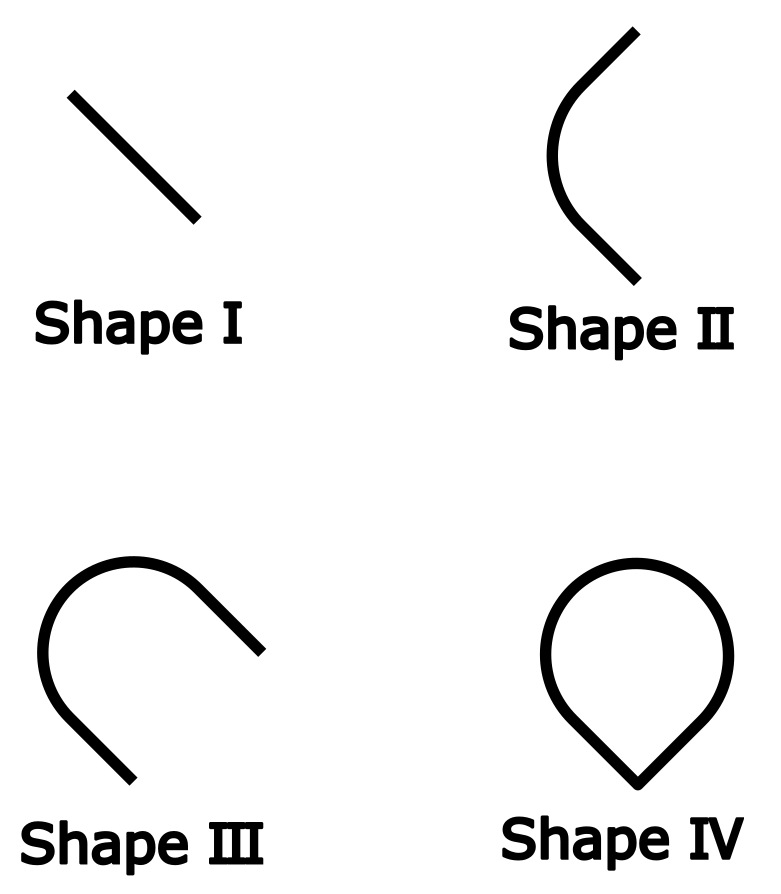}
    \caption{Shapes involved in single loop Kolams.}
\label{F11}
\end{figure}

\par Analyzing the dot Kolams under study, we assert that there are only 4 distinct shapes to be modeled. These shapes are shown in Figure \ref{F11}. These four basic Kolam shapes also cover distinct numbers of sides of the lattice cells: shape-I extends over only one side (distance of $L$), shape-II extends over two sides (distance equal to $2L$), shape-III extends over three sides (distance of $3L$) and shape-IV extends over four sides of the unit cell (distance of $4L$).  Most of the single-loop Kolams involve these 4 shapes only.\par

\begin{figure}
\centering
    \includegraphics[scale=1]{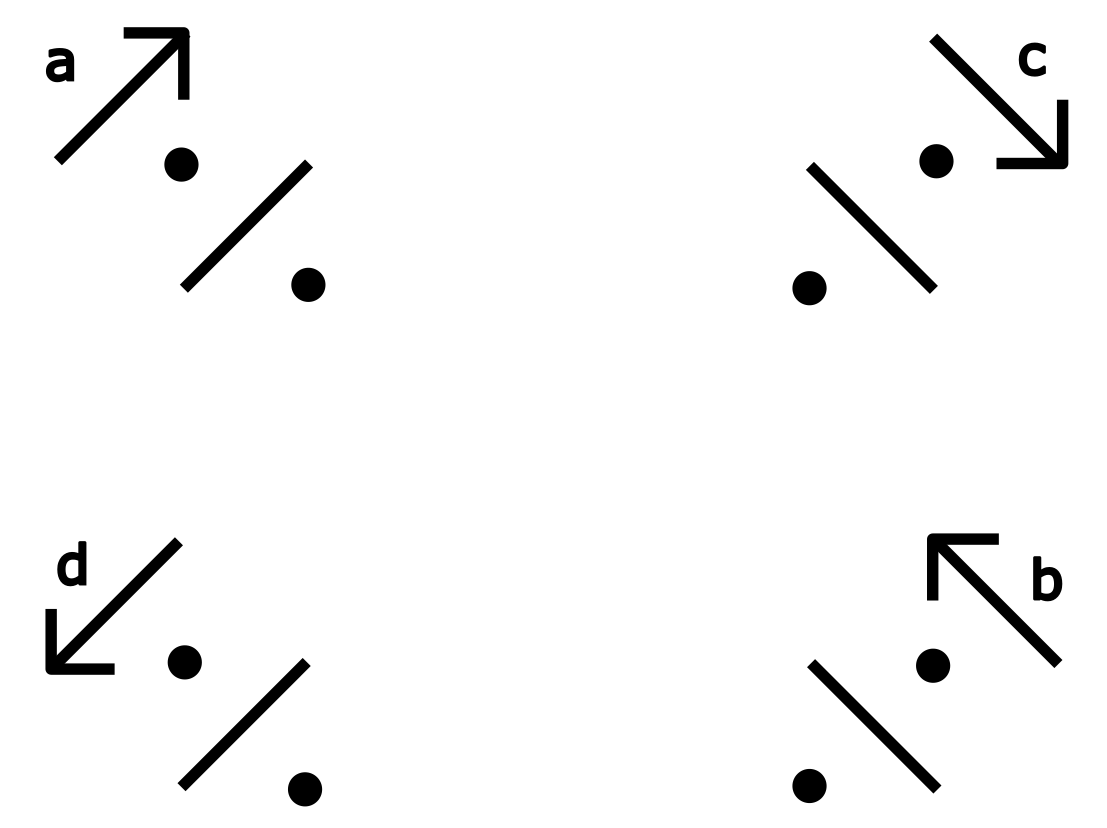}
    \caption{``$0$ changes in symbol'' shapes.}
\label{F12}
\end{figure}

\begin{figure}
\centering
    \includegraphics[scale=1]{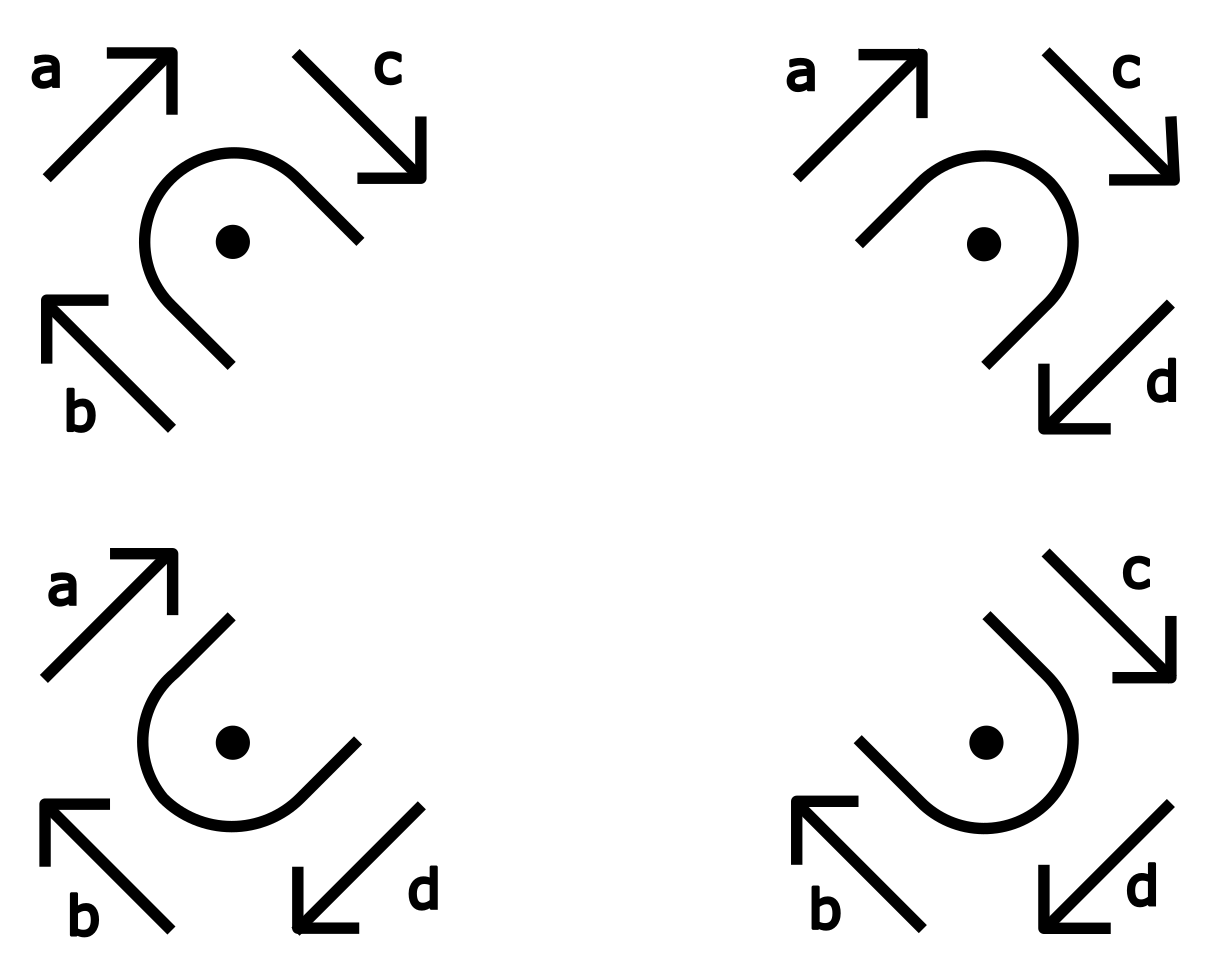}
    \caption{``$2$ changes in symbol'' shapes in clockwise direction.}
\label{F15}
\end{figure}

\begin{figure}
\centering
    \includegraphics[scale=1]{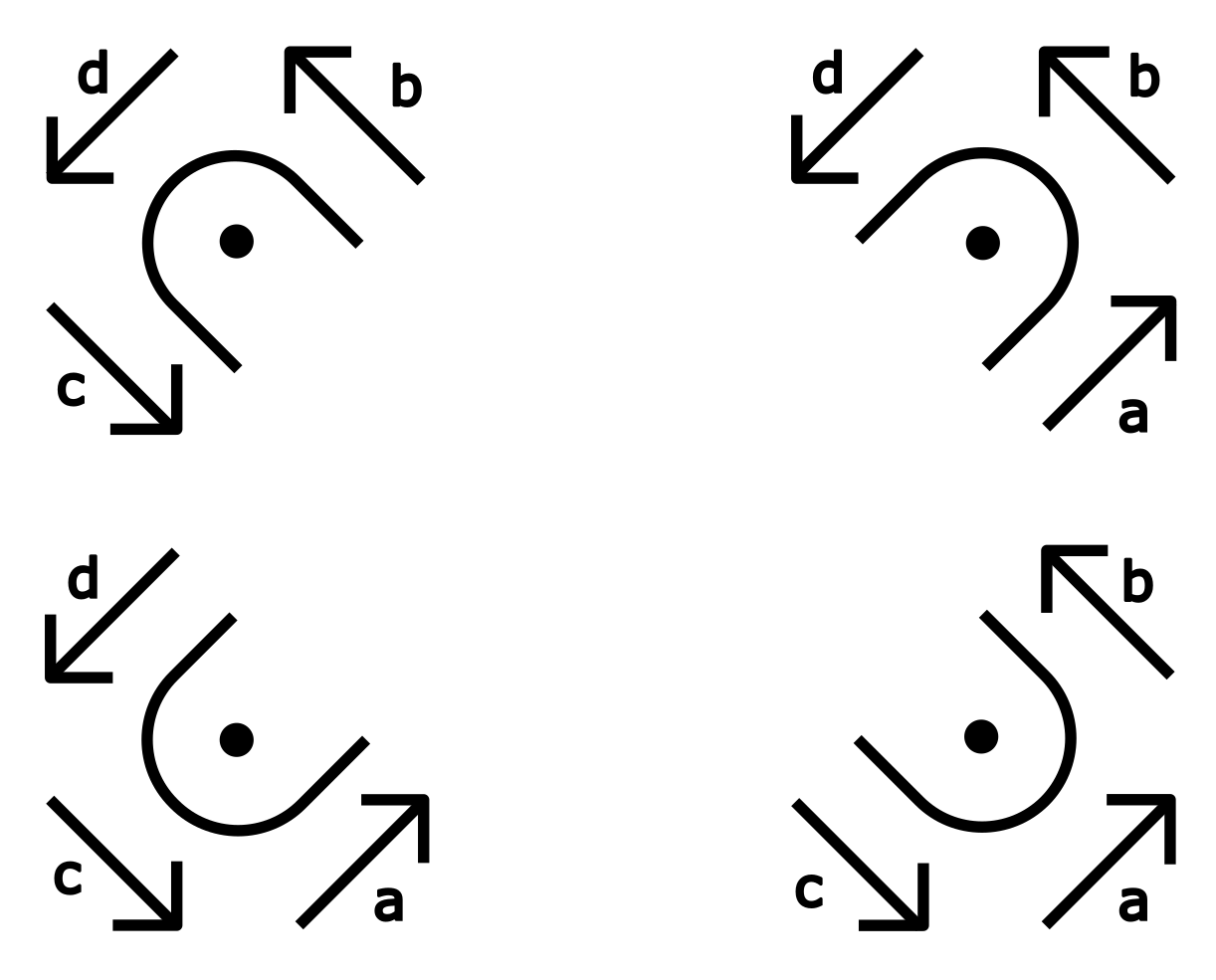}
    \caption{``$2$ changes in symbol'' shapes in anti-clockwise direction.}
\label{F16}
\end{figure}

\begin{figure}
\centering
    \includegraphics[scale=1]{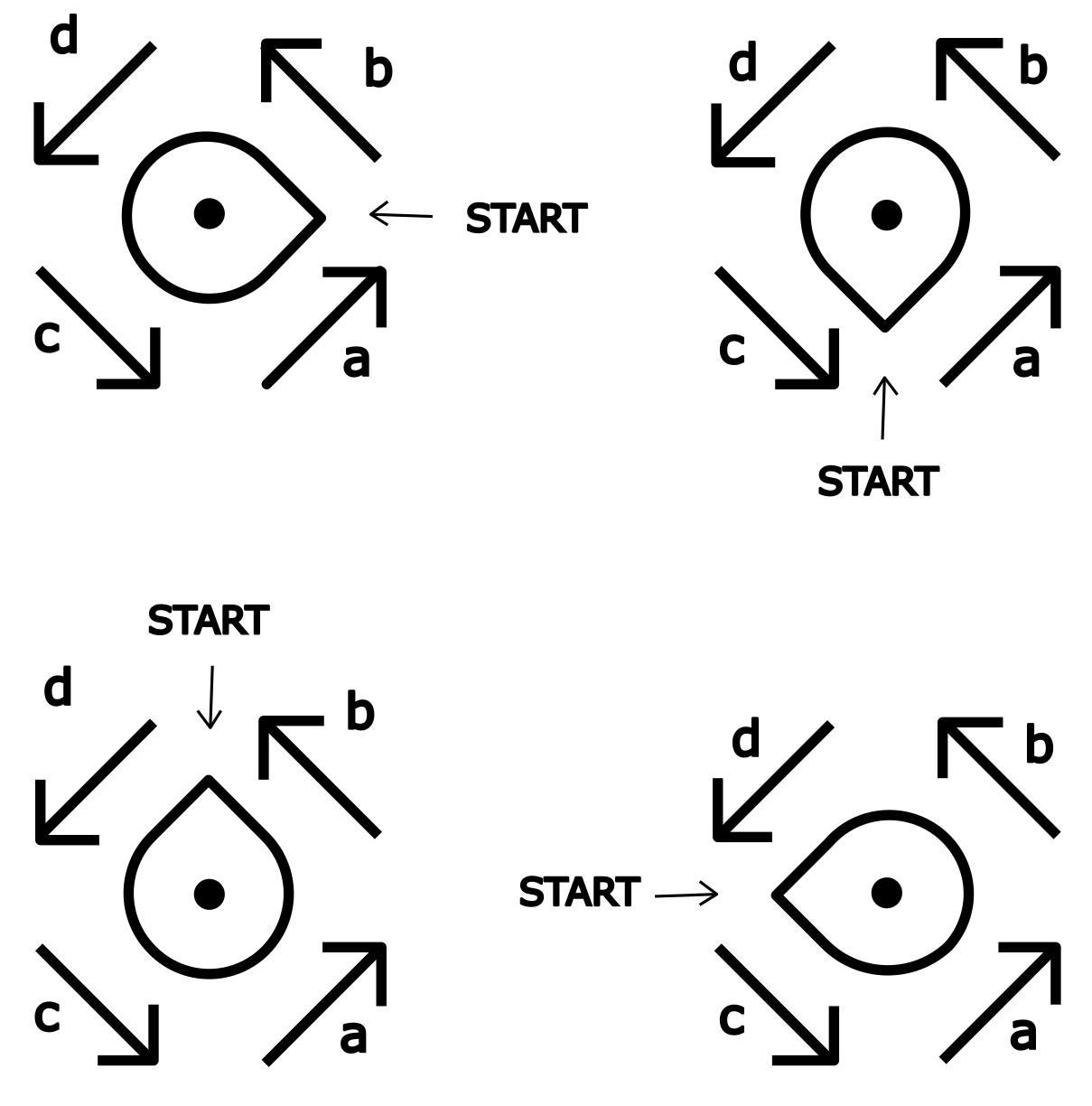}
    \caption{``$3$ changes in symbol'' shapes in anti-clockwise direction.}
\label{F17}
\end{figure}

Kolam drawing proceeds by shape determination after every distance of length $L$. Given the four distinct shapes possible, the shape determination for each segment  requires looking ahead to up to three successive steps (i.e. 4 successive elements of the sequence) starting from the current element, with a minimum of one step, which is adequate if there is no change in the direction (i.e. shape-I), to a maximum of 3 steps, if there is a direction change at every lattice point around the Kolam dot (i.e. shape-IV). The order of the changes in the angles determines how to draw a particular shape. Thus, these sequential patterns are nothing but the shape of a Kolam as it is drawn sequentially.  The pattern and its corresponding shape with an arrow mark showing its direction of drawing are given in Figures \ref{F12}-\ref{F18}. The encoded sequence pertaining to a Kolam is always processed from left to right. Therefore, the same Kolam can be represented by shifted versions or cyclic permutations of the encoding sequence. \par 

The shape determination conditions are modeled as a nested loop in our algorithm. We denote the elements of the sequence as follows: \par 
1. Current element($e_c$) - Outer loop element while iterating the inner loop\par
2. First element($e_1$) - First element succeeding the current element in the sequence\par
3. Second element($e_2$) - Second element succeeding the current element\par
4. Third element($e_3$) - Third element succeeding the current element\par
 
The four patterns correspond to the number of changes in the succeeding angles or number of changes in the succeeding elements of a sequence compared to the previous element of the sequence while iterating through it. They are: \par
Shape-I: $0$ consecutive changes $\equiv$ current element is the same as the first element. Example: ${aa}$ (look-ahead to 1 step) \par
Shape-II: $1$ consecutive change $\equiv$ current and first element are different AND first and second element are the same. Example: ${abb}$ (look ahead to 2 steps) \par
Shape-III: $2$ consecutive changes $\equiv$ current and first element are different AND first and second element are different AND second and third element are the same. Example: ${cabb}$ (look ahead to 3 steps) \par 
Shape-IV: $3$ consecutive changes $\equiv$ current and first element are different AND first and second element are different AND second and third element are different. Example:  ${dbac}$ (look ahead to 3 steps)\par

\subsection{Handling initialization issues}
\begin{figure}
\centering
    \includegraphics[scale=1]{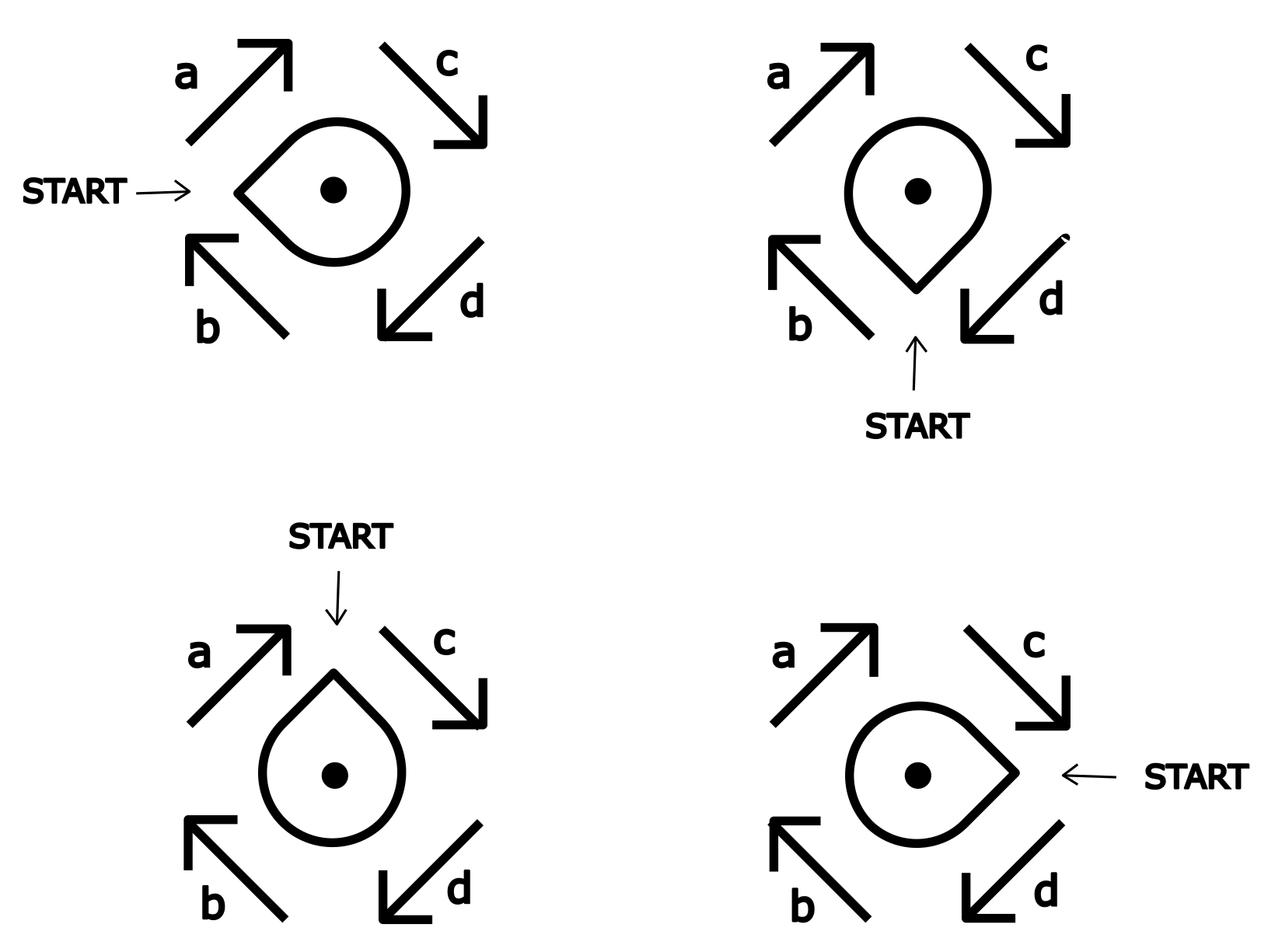}
    \caption{``$3$ changes in symbol'' shapes in clockwise direction.}
\label{F18}
\end{figure}

The Kolam sequence always begins as per the convention discussed in section \ref{Terminology}. But some Kolams begin with the shape-IV(${dbac}$) given in Figure \ref{F18} which starts drawing at the right lattice point of the beginning dot. This location is not according to our convention of beginning of a sequence. This situation is covered by the following rule incorporated into the present algorithm: If the third element from the last of the sequence is `$d$', then `$d$' will be added in the beginning of the sequence before iterating the sequence and the shape-IV(${dbac}$) will be drawn first when we start iterating the sequence. If the third element from the last of the sequence is `$b$', then `$b$' will be added in the beginning of the sequence and also at the third position from the last of the sequence before iterating the sequence. It indicates that the Kolam is not beginning with the shape-IV(${dbac}$) and helps in identifying the final shape.    \par

\subsection{Terminating condition}
\begin{figure}
\centering
    \includegraphics[scale=1]{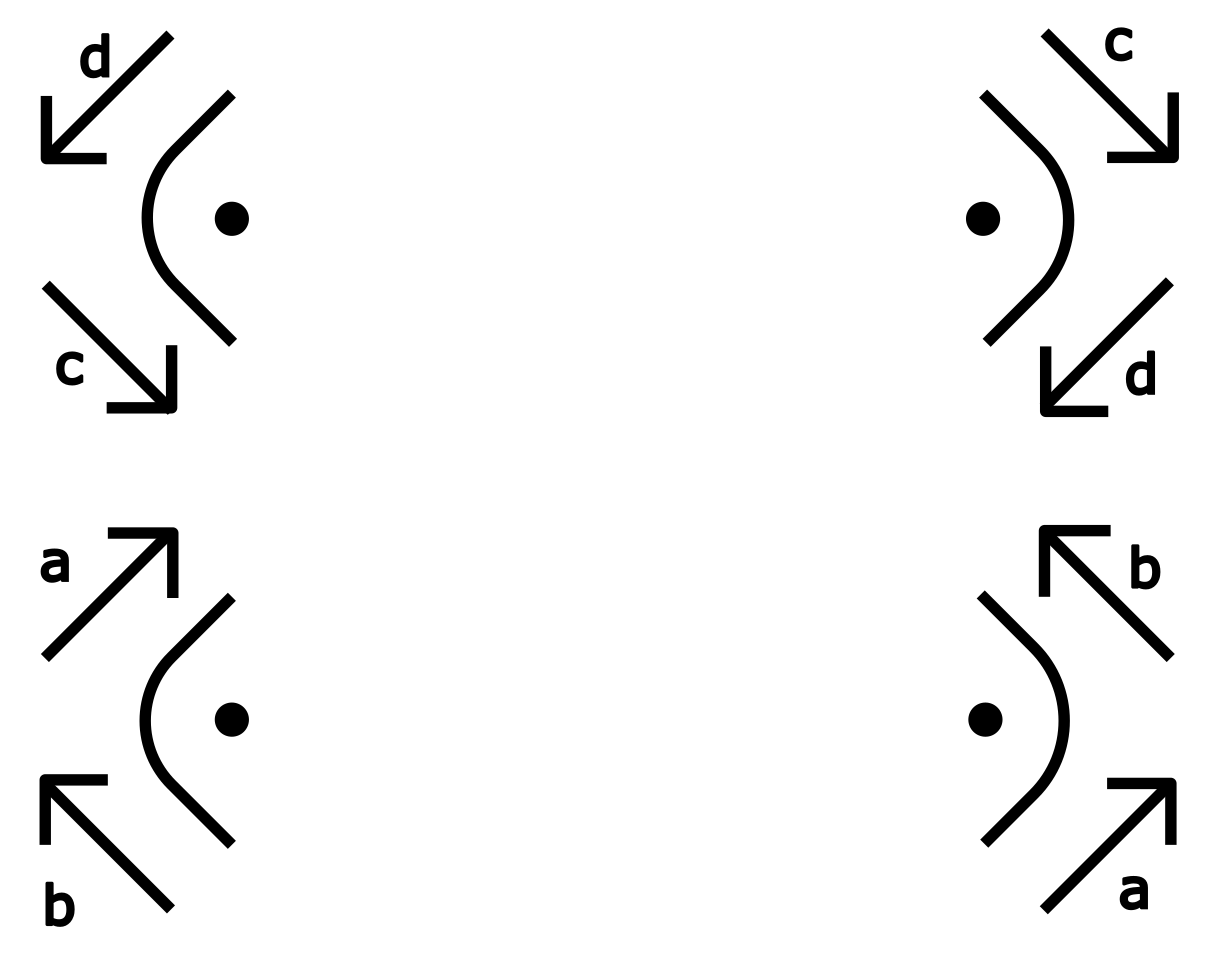}
    \caption{``$1$ change in symbol'' shapes in vertical direction.}
\label{F13}
\end{figure}

\begin{figure}
\centering
    \includegraphics[scale=1]{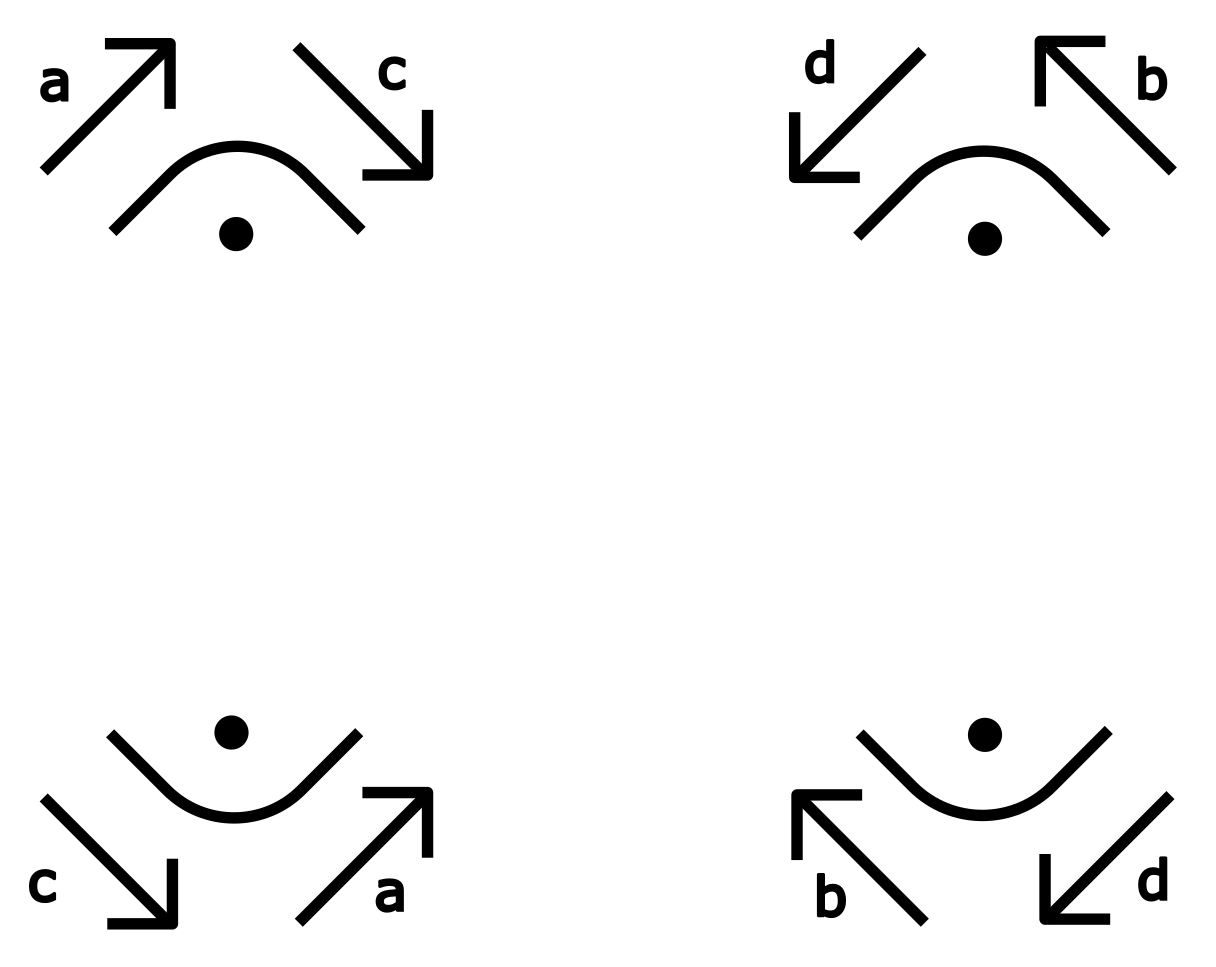}
    \caption{``$1$ change in symbol'' shapes in horizontal direction.}
\label{F14}
\end{figure}

Our algorithmic conditions look ahead to the maximum 3 elements from the current element to determine the shape to be drawn. Two $z$'s are added at the end of the sequence which act as nulls (symbols with no effect) if the final shape is shape-I or shape-II, otherwise it leads to an error or a wrong shape selection. For example, if the sequence ends with `$dzz$', there is a change from $d$$\rightarrow$$z$ and no change from $z$$\rightarrow$$z$. Though it looks like a shape-II pattern, `$z$' is a null and it is not any possible pattern of shape-II specified in Figures \ref{F13} and \ref{F14}. As it does not meet any of our algorithmic conditions, the loop will be terminated after iterating through these $z$'s. The Kolam drawing will be finished. For shape-III and shape-IV, even if there are no $z$'s, no problem will arise. For example, if the sequence ends with `$dcab$', the shape-IV ($dcab$) will be drawn and the loop will be terminated without any errors. The Kolam drawing will be finished. But if $z$'s are not added, it raises an error or selects a wrong shape when the final shape is either shape-I or shape-II. For example, if the sequence ends with `$aab$', there is no change from $a$$\rightarrow$$a$ and this is a shape-I ($a$) pattern but there is a change from $a$$\rightarrow$$b$. It is the first condition with no shape specified in section \ref{Kolam Simulation Algorithm}. Now `$a$' becomes the current element. There is a change from $a$$\rightarrow$$b$ and it needs to take second step to determine the shape. As there is no element after `$b$', it will raise an error. To avoid these errors and discrepancies and to encode all Kolams in a standard way, we add $z$'s in the end of a Kolam sequence. The loop will always be terminated after iterating through these $z$'s. \par

\begin{figure}
\centering
    \includegraphics[scale=0.6]{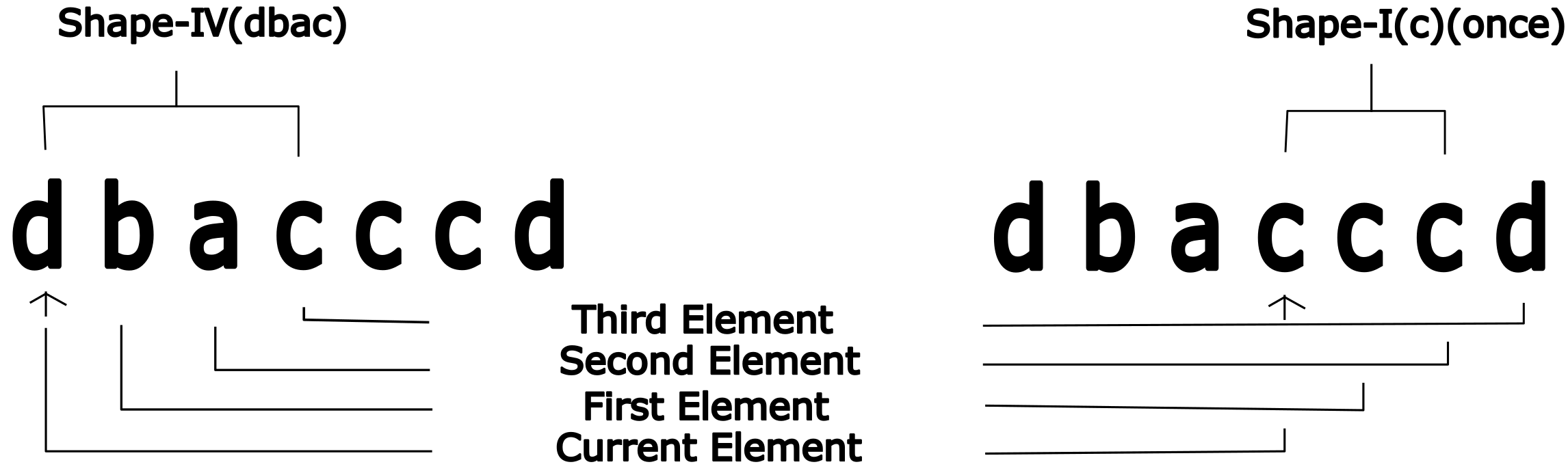}
    \caption{Iterating through the sequence: an example.}
\label{F19}
\end{figure}

\section{Kolam Simulation Algorithm}\label{Kolam Simulation Algorithm}
The algorithm for Kolam simulation depends on the number of changes in the successive symbols, which gives us the pattern for drawing one of the 4 possible shapes discussed in section \ref{Data Pre-processing}. The sequence is iterated and while it is being iterated, it checks for the following conditions.\par
1. $n$ consecutive no change (the same symbols or $0$ changes) – draw the shape-I $n-1$ times and its corresponding orientation (from memory while iterating). If $n=1$, then no shape will be drawn and it is a no-shape condition.\par
2. $1$ consecutive change or $2$ consecutive changes with the second element equal to `$z$' – draw the shape-II and its corresponding orientation (from memory while iterating).\par
3. $2$ consecutive changes and the second element is not equal to `$z$' – draw the shape-III and its corresponding orientation (from memory while iterating).\par
4. $3$ consecutive changes and the third element is not equal to `$z$' – draw the shape-IV and its corresponding orientation (from memory while iterating).\par

Formally, a complete description of the algorithm is described in Algorithm~\ref{alg:MAIN}.

\begin{algorithm}
\caption{Kolam Simulation Algorithm}\label{alg:MAIN}
\begin{algorithmic}
\Procedure{Kolam Simulation}{$list$}  

\State \textbf{Input:} $list = x_1, x_2, \ldots, x_n, z, z$
\State \textbf{Output:} Kolam

\State $i = 0$
\State $j = 0$
\While{$i < (len(list)-3)$}
\If{$list_{i+3} \neq z  \land list_{i} \neq list_{i+1} \land list_{i+1} \neq list_{i+2} \land list_{i+2} \neq list_{i+3}$}
    \State $e_c \gets list_i$
    \State $e_1 \gets list_{i+1}$
    \State $e_2 \gets list_{i+2}$
    \State $e_3 \gets list_{i+3}$
    \State SHAPE IV ($e_c$, $e_1$, $e_2$, $e_3$)    \Comment{see Fig. \ref{F17} and \ref{F18}}
    \State $i \gets i + 3$
\ElsIf{$list_{i+2} \neq z \land list_{i} \neq list_{i+1} \land list_{i+1} \neq list_{i+2}$}
    \State $e_c \gets list_i$
    \State $e_1 \gets list_{i+1}$
    \State $e_2 \gets list_{i+2}$
    \State SHAPE III ($e_c$, $e_1$, $e_2$)    \Comment{see Fig. \ref{F15} and \ref{F16}}
    \State $i \gets i + 2$
\ElsIf{$list_{i} \neq list_{i+1}$}
    \State $e_c \gets list_i$
    \State $e_1 \gets list_{i+1}$
    \State SHAPE II ($e_c$, $e_1$)  \Comment{see Fig. \ref{F13} and \ref{F14}}
    \State $i \gets i + 1$
\ElsIf{$list_{i} = list_{i + 1}$}
    \State SHAPE I ($list$, $i$)       \Comment{see Fig. \ref{F12}}
    \State $i \gets i + j$
\EndIf
\EndWhile

\EndProcedure
\end{algorithmic}
\end{algorithm}

\begin{algorithm}
\caption{Shape I}\label{alg:SHAPE I}
\begin{algorithmic}
\Procedure{Shape I}{$list$, $i$}  

\State \textbf{Input:} $list$, $i$
\State \textbf{Output:} Kolam Shape I

\State{$j = 0$}
\While{$j < (len(list))$}
\If{$list_{i+j} = list_{i+j+1}$}
\State $j \gets j + 1$
\Else
\State break
\EndIf

\State $k \gets 1$
\State $e_c \gets list_i$
\EndWhile

\While{$k < j$}            
\If{$e_c = a$}
\State Draw Shape I (a) \Comment{see Fig. \ref{F12}}
\State $k \gets k + 1$
\ElsIf{$e_c = b$}
\State Draw Shape I (b) \Comment{see Fig. \ref{F12}}
\State $k \gets k + 1$
\ElsIf{$e_c = c$}
\State Draw Shape I (c) \Comment{see Fig. \ref{F12}}
\State $k \gets k + 1$
\ElsIf{$e_c = d$}
\State Draw Shape I (d) \Comment{see Fig. \ref{F12}}
\State $k \gets k + 1$
\EndIf
\EndWhile

\EndProcedure
\end{algorithmic}
\end{algorithm}

\begin{algorithm}
\caption{Shape II}\label{alg:SHAPE II}
\begin{algorithmic}
\Procedure{Shape II}{$e_c$, $e_1$} 

\State Input: $e_c$, $e_1$
\State Output: Kolam Shape II

\If{$e_c = b \land e_1 = d$}
\State Draw Shape II (bd)   \Comment{see Fig. \ref{F14}}
\ElsIf{$e_c = d \land e_1 = c$}  
\State Draw Shape II (dc)   \Comment{see Fig. \ref{F13}}
\ElsIf{$e_c = c \land e_1 = a$} 
\State Draw Shape II (ca)   \Comment{see Fig. \ref{F14}}
\ElsIf{$e_c = a \land e_1 = b$}
\State Draw Shape II (ab)   \Comment{see Fig. \ref{F13}}    
\ElsIf{$e_c = a \land e_1 = c$}        
\State Draw Shape II (ac)   \Comment{see Fig. \ref{F14}}
\ElsIf{$e_c = c \land e_1 = d$}
\State Draw Shape II (cd)    \Comment{see Fig. \ref{F13}}       
\ElsIf{$e_c = d \land e_1 = b$}            
\State Draw Shape II (db)   \Comment{see Fig. \ref{F14}}
\ElsIf{$e_c = b \land e_1 = a$}
\State Draw Shape II (ba)   \Comment{see Fig. \ref{F13}}
\EndIf

\EndProcedure
\end{algorithmic}
\end{algorithm}

\begin{algorithm}
\caption{Shape III}\label{alg:SHAPE III}
\begin{algorithmic}
\Procedure{Shape III}{$e_c$, $e_1$, $e_2$}  

\State Input: $e_c$, $e_1$, $e_2$
\State Output: Kolam Shape III

\If{$e_c = b \land e_1 = d \land e_2 = c$}
\State Draw Shape III (bdc)     \Comment{see Fig. \ref{F16}}
\ElsIf{$e_c = d \land e_1 = c \land e_2 = a$}
\State Draw Shape III (dca)     \Comment{see Fig. \ref{F16}}
\ElsIf{$e_c = c \land e_1 = a \land e_2 = b$}
\State Draw Shape III (cab)     \Comment{see Fig. \ref{F16}}
\ElsIf{$e_c = a \land e_1 = b \land e_2 = d$}
\State Draw Shape III (abd)     \Comment{see Fig. \ref{F16}}
\ElsIf{$e_c = b \land e_1 = a \land e_2 = c$}
\State Draw Shape III (bac)     \Comment{see Fig. \ref{F15}}    
\ElsIf{$e_c = a \land e_1 = c \land e_2 = d$}            
\State Draw Shape III (acd)     \Comment{see Fig. \ref{F15}}
\ElsIf{$e_c = c \land e_1 = d \land e_2 = b$}
\State Draw Shape III (cdb)     \Comment{see Fig. \ref{F15}}
\ElsIf{$e_c = d \land e_1 = b \land e_2 = a$}
\State Draw Shape III (dba)     \Comment{see Fig. \ref{F15}}
\EndIf

\EndProcedure
\end{algorithmic}
\end{algorithm}

\begin{algorithm}
\caption{Shape IV}\label{alg:SHAPE IV}
\begin{algorithmic}
\Procedure{Shape IV}{$e_c$, $e_1$, $e_2$, $e_3$}

\State Input: $e_c$, $e_1$, $e_2$, $e_3$
\State Output: Kolam Shape IV

\If{$e_c = a \land e_1 = b \land e_2 = d \land e_3 = c$}
\State Draw Shape IV (abdc)         \Comment{see Fig. \ref{F17}}
\ElsIf{$e_c = b \land e_1 = d \land e_2 = c \land e_3 = a$} 
\State Draw Shape IV (bdca)         \Comment{see Fig. \ref{F17}}
\ElsIf{$e_c = d \land e_1 = c \land e_2 = a \land e_3 = b$}            
\State Draw Shape IV (dcab)         \Comment{see Fig. \ref{F17}}        
\ElsIf{$e_c = c \land e_1 = a \land e_2 = b \land e_3 = d$} 
\State Draw Shape IV (cabd)         \Comment{see Fig. \ref{F17}}
\ElsIf{$e_c = b \land e_1 = a \land e_2 = c \land e_3 = d$}
\State Draw Shape IV (bacd)         \Comment{see Fig. \ref{F18}}
\ElsIf{$e_c = a \land e_1 = c \land e_2 = d \land e_3 = b$}
\State Draw Shape IV (acdb)         \Comment{see Fig. \ref{F18}}    
\ElsIf{$e_c = c \land e_1 = d \land e_2 = b \land e_3 = a$}
\State Draw Shape IV (cdba)         \Comment{see Fig. \ref{F18}}         
\ElsIf{$e_c = d \land e_1 = b \land e_2 = a \land e_3 = c$}        
\State Draw Shape IV (dbac)         \Comment{see Fig. \ref{F18}}        
\EndIf

\EndProcedure
\end{algorithmic}
\end{algorithm}

We can also use the sequence in a cyclic iteration without $z$'s and adding `$b$' in the third position from the last of the sequence. In this case it will draw the Kolam infinite times without stopping. The output of this Kolam is given in Figure \ref{F29}.

\begin{table}
\centering
\caption{Navagraha Kolams and corresponding digital sequences}
\vspace{0.1in}
\begin{tabular}{@{}ll@{}}

\toprule
\multicolumn{2}{c}{Lattice angle sequences}                                                                                                                                     \\ \midrule
Kolam 1 & \begin{tabular}[c]{@{}l@{}}(b, a, c, c, c, d, d, c, a, b, b, a, a, a, c, d, d, c, c, d, b, a, a, b, b, b, d, d, d, c, c, d, b, a, a, b, z, z)\end{tabular}                     \\ \midrule
Kolam 2 & \begin{tabular}[c]{@{}l@{}}(b, a, c, c, c, c, c, d, b, a, a, b, d, d, d, c, a, b, b, b, d, c, c, d, b, a, a, a, a, a, c, d, b, b, d, d, z, z)\end{tabular}                     \\ \midrule
Kolam 3 & \begin{tabular}[c]{@{}l@{}}(b, a, c, d, d, c, c, d, b, a, a, a, a, b, d, c, c, c, c, d, b, a, a, b, b, a, c, d, d, d, d, c, a, b, b, b, z, z)\end{tabular}                     \\ \midrule
Kolam 4 & \begin{tabular}[c]{@{}l@{}}(b, a, c, c, c, d, d, d, b, a, c, c, a, a, c, d, b, b, b, a, a, a, c, d, d, c, a, b, b, b, d, d, d, c, a, b, z, z)\end{tabular}                     \\ \midrule
Kolam 5 & \begin{tabular}[c]{@{}l@{}}(b, a, c, c, c, c, c, d, b, a, a, b, b, b, d, d, d, c, c, d, b, a, a, a, a, a, c, d, d, d, d, c, a, b, b, b, z, z)\end{tabular}                     \\ \midrule
Kolam 6 & \begin{tabular}[c]{@{}l@{}}(b, a, c, c, c, c, a, b, b, b, d, d, d, c, a, a, a, a, c, d, d, d, d, d, b, a, c, c, a, a, c, d, b, b, b, b, z, z)\end{tabular}                     \\ \midrule
Kolam 7 & \begin{tabular}[c]{@{}l@{}}(b, a, c, d, d, c, c, c, a, b, d, d, b, a, a, a, c, c, c, d, b, a, a, b, b, b, d, c, a, a, c, d, d, d, b, b, z, z)\end{tabular}                     \\ \midrule
Kolam 8 & \begin{tabular}[c]{@{}l@{}}(b, a, c, c, a, a, c, d, d, d, b, b, d, c, c, c, a, a, c, d, b, b, d, d, b, a, a, a, c, c, a, b, b, b, d, d, z, z)\end{tabular}                     \\ \midrule
Kolam 9 & \multicolumn{1}{c}{\begin{tabular}[c]{@{}c@{}}(b, a, c, d, d, c, a, a, a, a, c, d, b, b, d, c, c, c, c, d, b, a, a, b, d, d, d, d, b, a, c, c, a, b, b, b, z, z)\end{tabular}} \\ \bottomrule
\end{tabular}
\label{T2}
\end{table}

\begin{table}
\centering
\caption{Rhombic Kolams and corresponding digital sequences}
\vspace{0.1in}
\begin{tabular}{@{}ll@{}}
\toprule
\multicolumn{2}{c}{Lattice angle sequences}                                                                                                                                                                                                                                                                                                                           \\ \midrule
1-3-1 Kolam & (b, a, c, c, c, d, b, a, a, a, c, d, b, b, b, a, c, d, d, d, z, z)                                                                                                                                                                                                                                                                                                  \\ \midrule
1-5-1 Kolam & \begin{tabular}[c]{@{}l@{}}(b, a, c, c, c, d, b, a, a, a, a, a, c, d, b, b, b, a, c, d, d, d, b, a, c, c, c, d, d, c, c, d, b, a, a, a, a, a, c, d, \\ b, b, d, c, c, d, b, b, b, b, d, d, z, z)\end{tabular}                                                                                                                                                       \\ \midrule
1-7-1 Kolam & \begin{tabular}[c]{@{}l@{}}(b, a, c, c, c, c, a, b, b, b, b, a, c, c, c, c, c, d, d, d, b, a, c, c, c, d, b, a, a, a, c, d, b, b, b, a, a, a, a, a, \\ c, d, d, d, d, c, a, a, a, a, c, d, b, b, b, b, d, c, c, c, c, d, b, b, b, b, b, a, a, a, c, d, b, b, b, a, c, d, d, d, \\ b, a, c, c, c, d, d, d, d, d, b, a, a, a, a, b, d, d, d, d, z, z)\end{tabular} \\ \bottomrule
\end{tabular}
\label{T3}
\end{table}

\section{Results and Discussion}
\begin{figure}
\centering
    \includegraphics[scale=0.5]{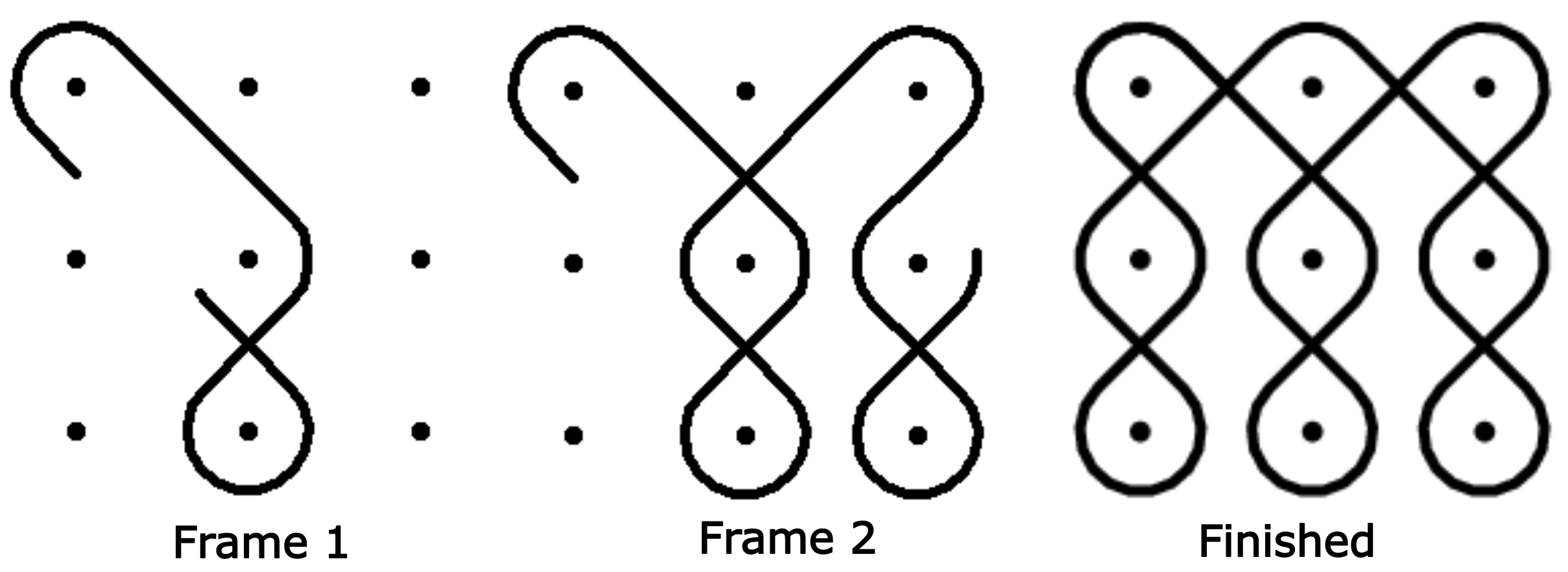}
    \caption{Navagraha ($3\times3$ square) Kolam $1$.}
\label{F20}
\end{figure}

\begin{figure}
\centering
    \includegraphics[scale=0.5]{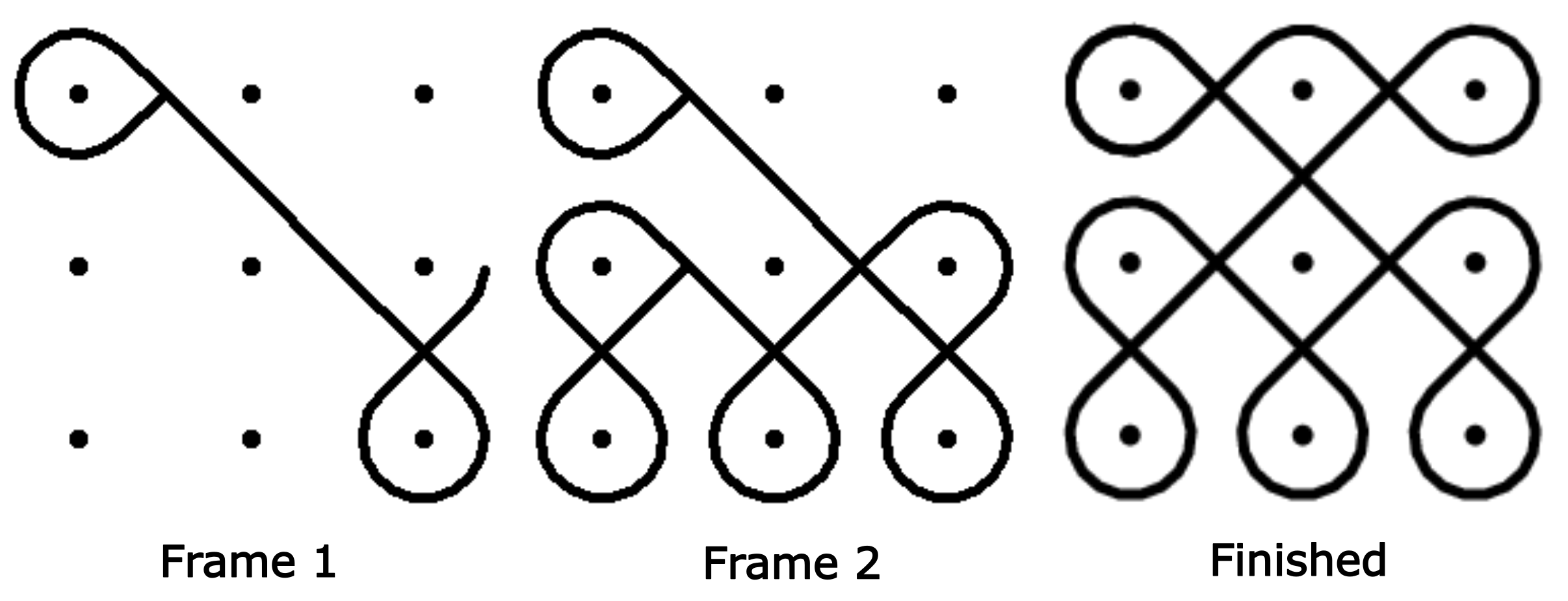}
    \caption{Navagraha ($3\times3$ square) Kolam $2$.}
\label{F21}
\end{figure}

\begin{figure}
\centering
    \includegraphics[scale=0.5]{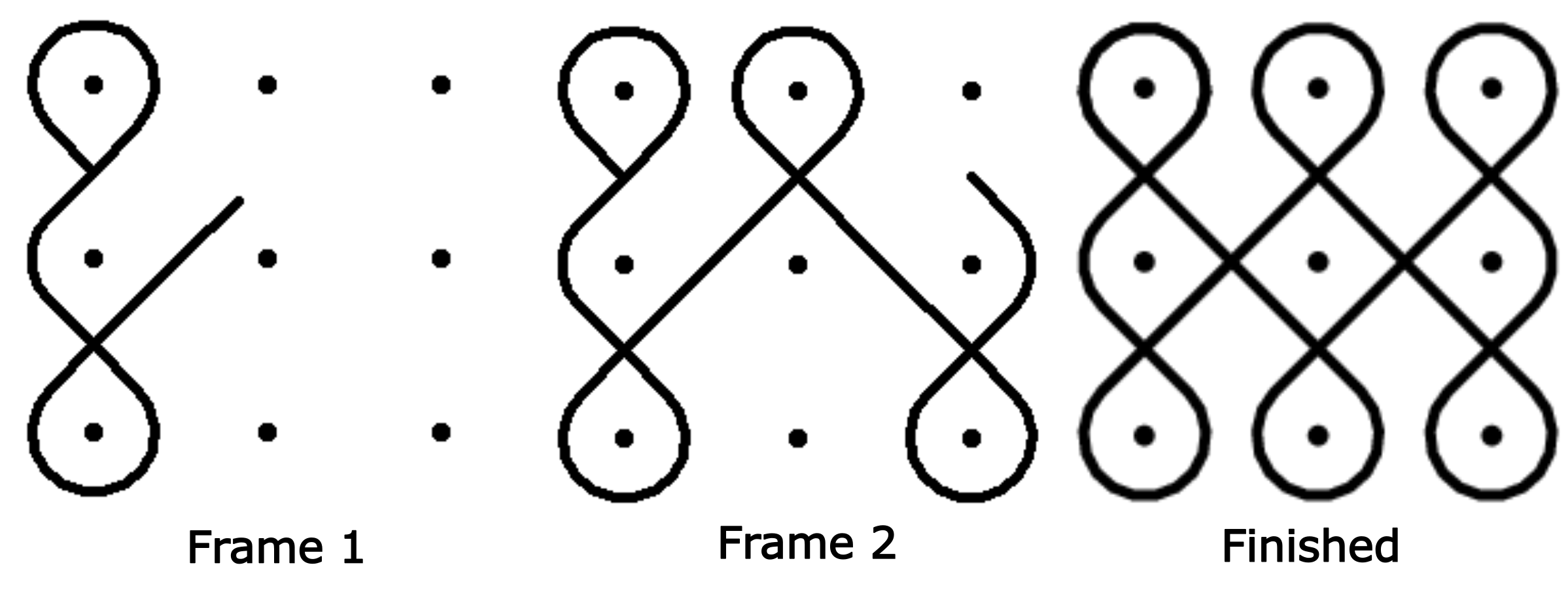}
    \caption{Navagraha ($3\times3$ square) Kolam $3$.}
\label{F22}
\end{figure}

\begin{figure}
\centering
    \includegraphics[scale=0.5]{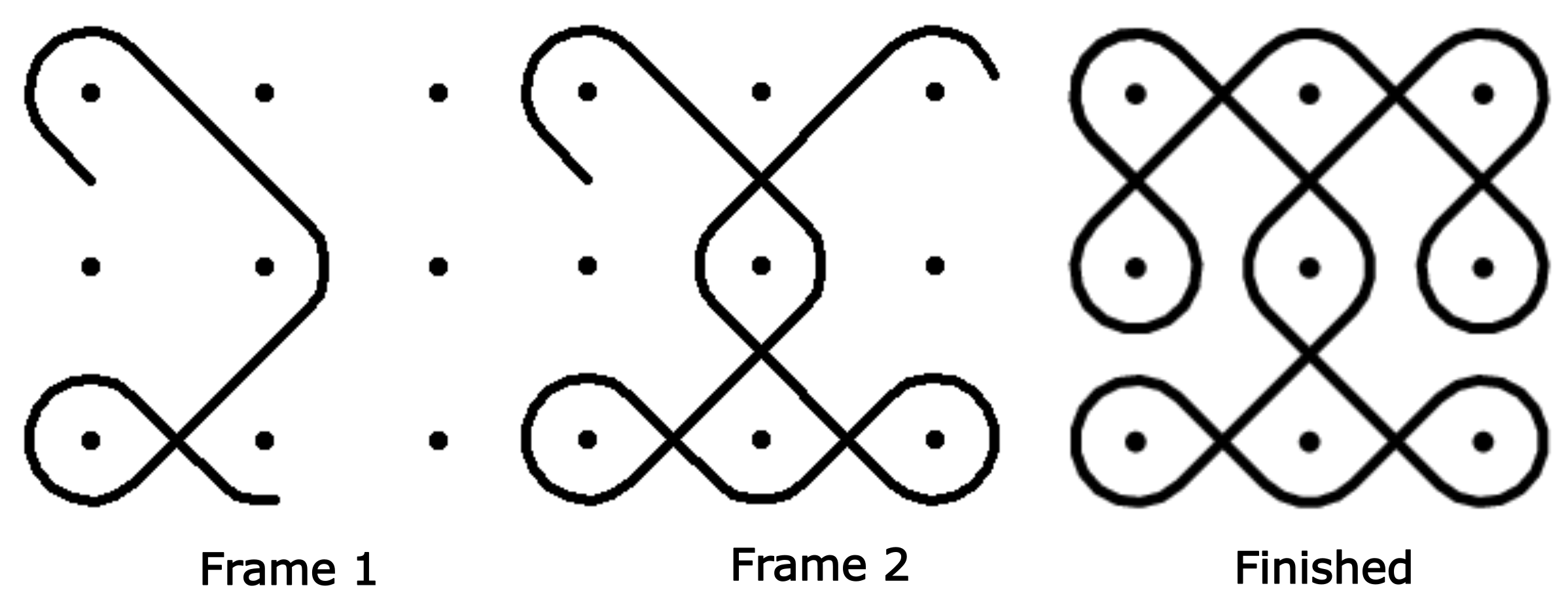}
    \caption{Navagraha ($3\times3$ square) Kolam $4$.}
\label{F23}
\end{figure}

\begin{figure}
\centering
    \includegraphics[scale=0.5]{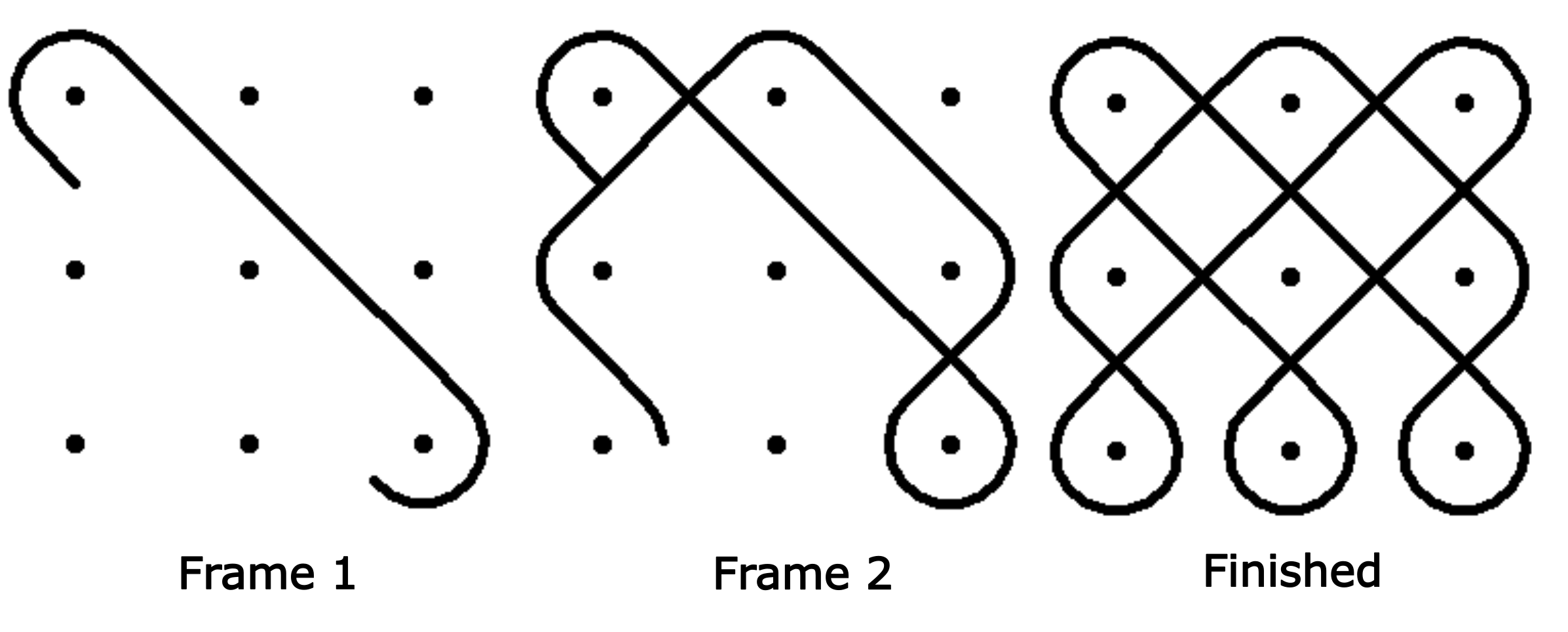}
    \caption{Navagraha ($3\times3$ square) Kolam $5$.}
\label{F24}
\end{figure}

\begin{figure}
\centering
    \includegraphics[scale=0.5]{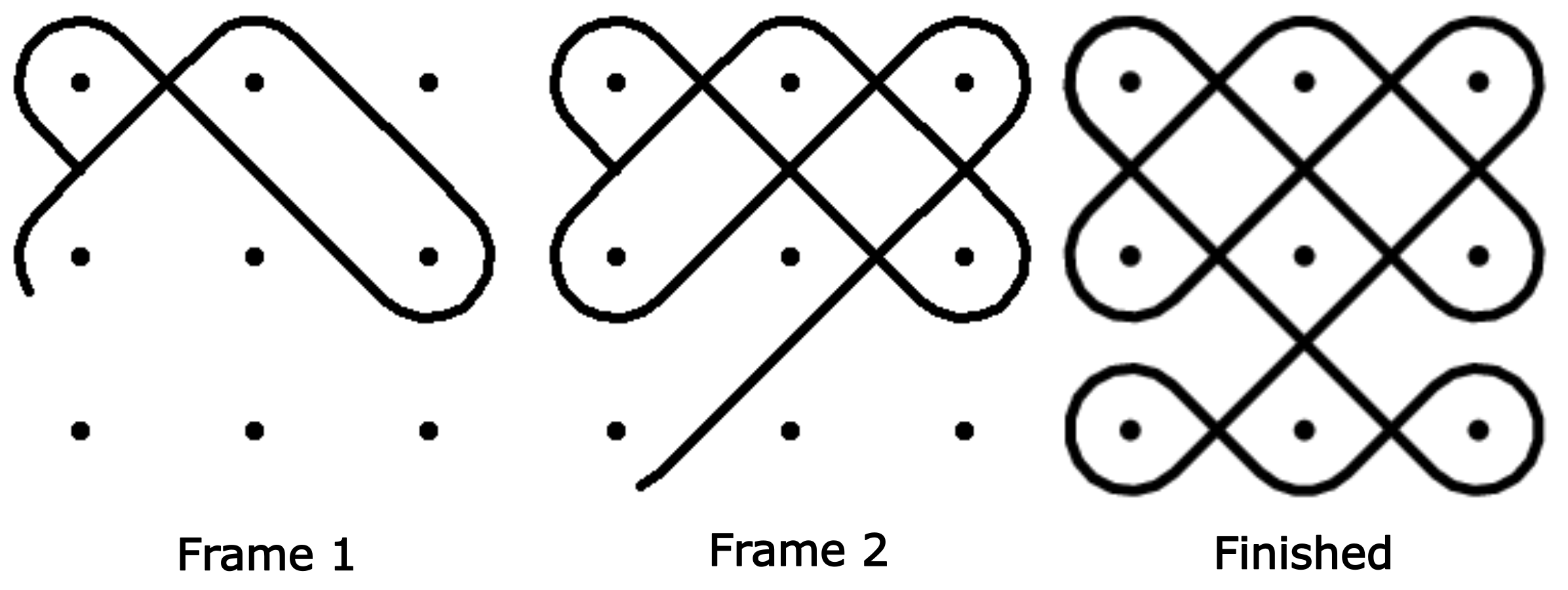}
    \caption{Navagraha ($3\times3$ square) Kolam $6$.}
\label{F25}
\end{figure}

\begin{figure}
\centering
    \includegraphics[scale=0.5]{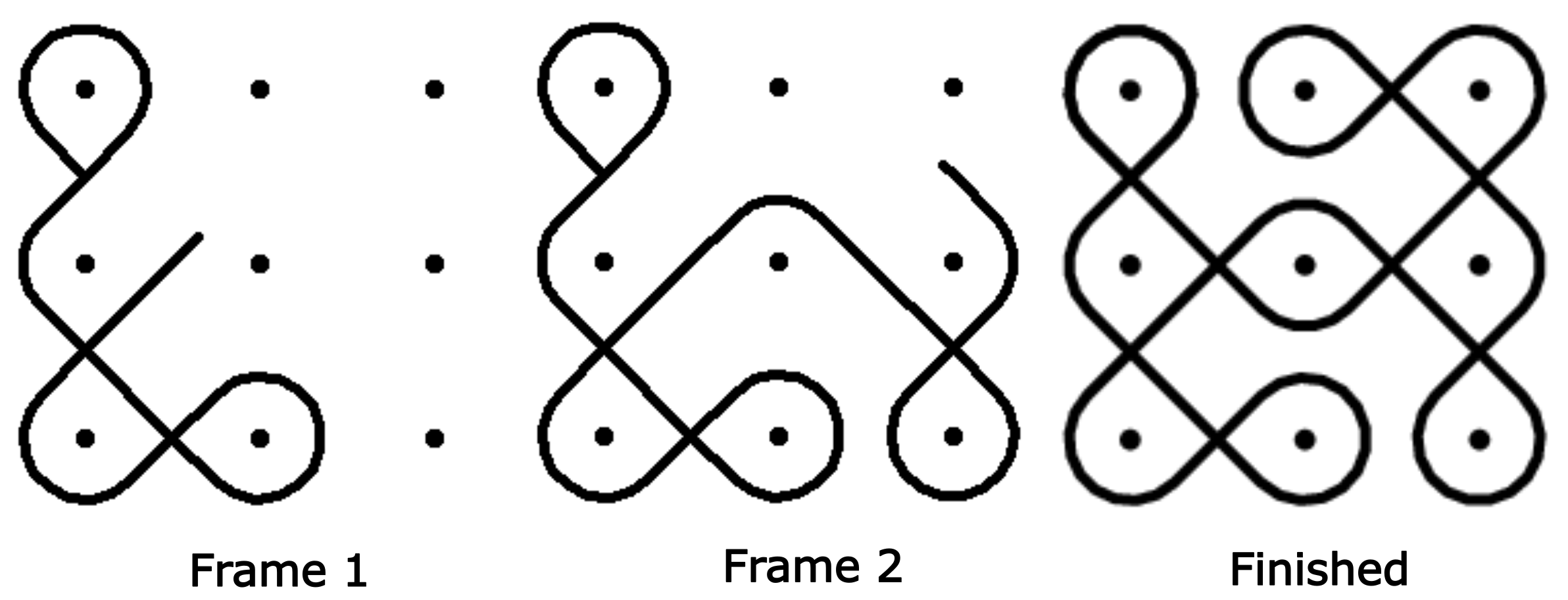}
    \caption{Navagraha ($3\times3$ square) Kolam $7$.}
\label{F26}
\end{figure}

\begin{figure}
\centering
    \includegraphics[scale=0.5]{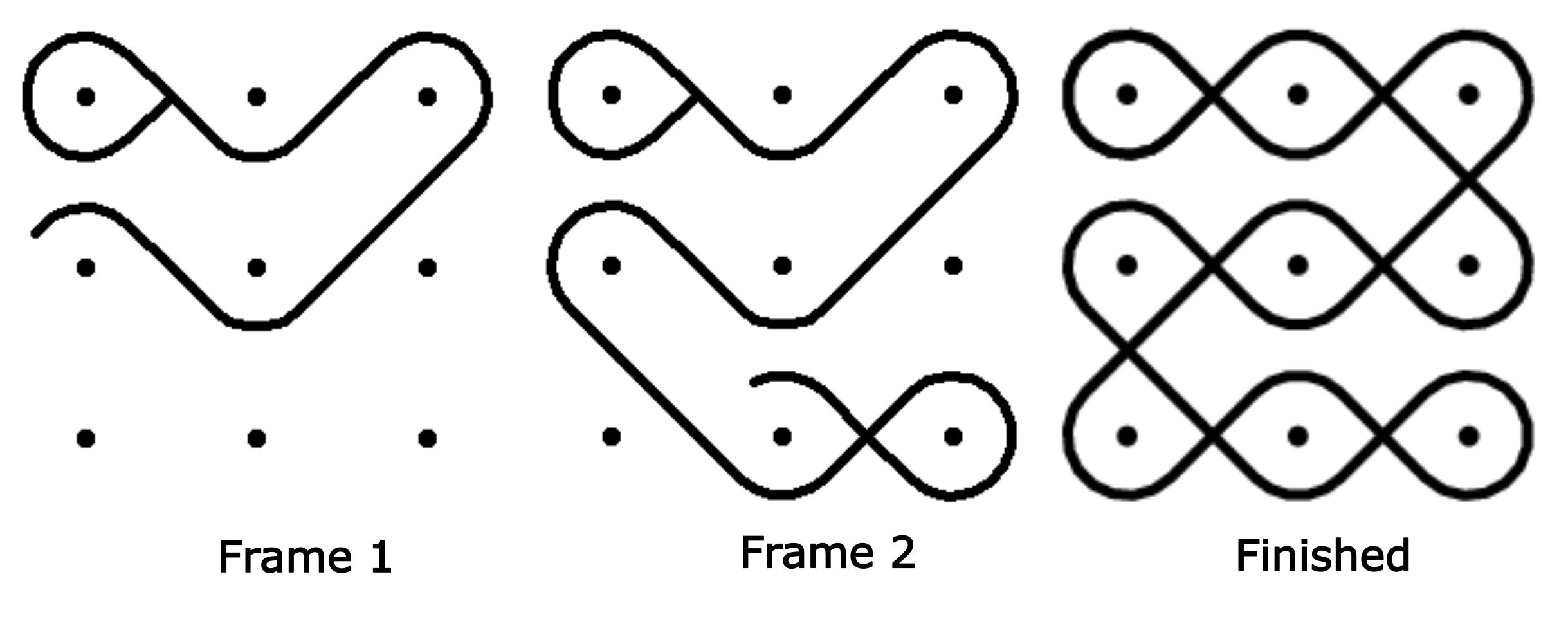}
    \caption{Navagraha ($3\times3$ square) Kolam $8$.}
\label{F27}
\end{figure}

\begin{figure}
\centering
    \includegraphics[scale=0.5]{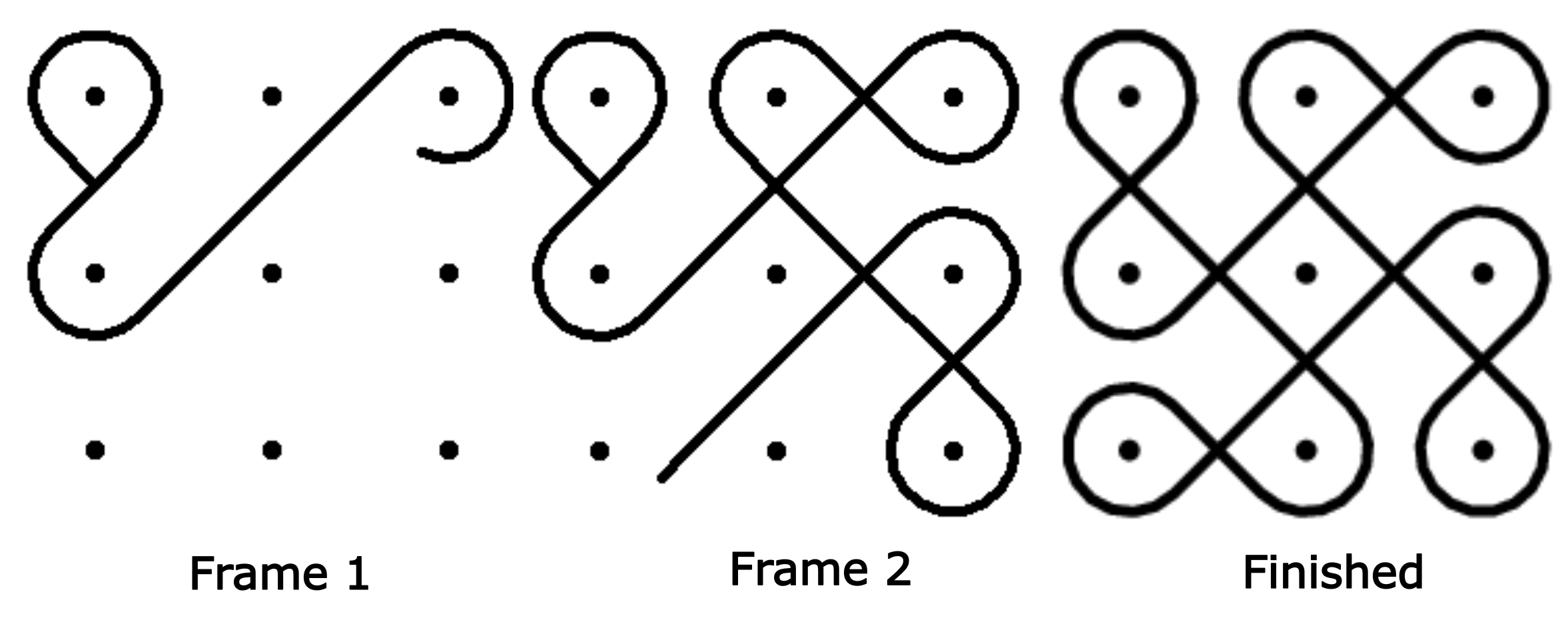}
    \caption{Navagraha ($3\times3$ square) Kolam $9$.}
\label{F28}
\end{figure}

\begin{figure}
\centering
    \includegraphics[scale=0.5]{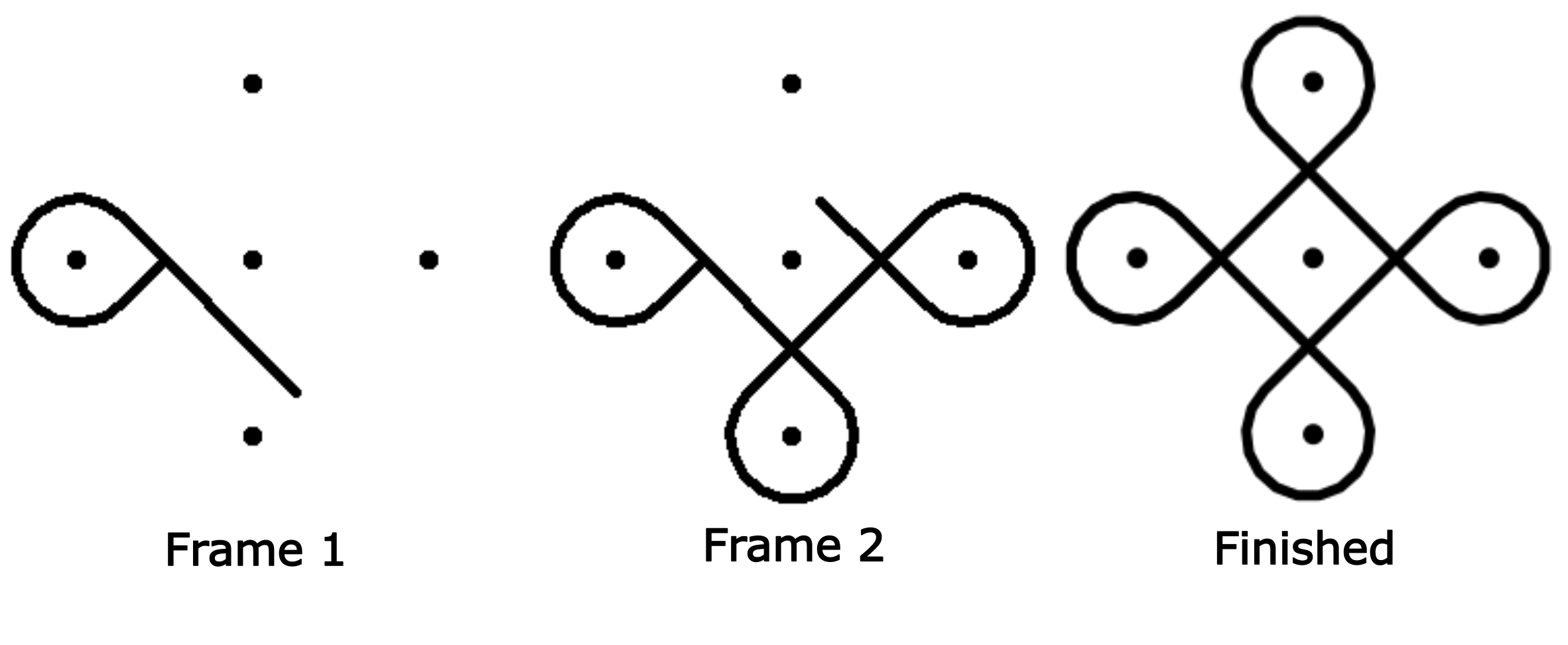}
    \caption{A $1-3-1$ rhombic Kolam.}
\label{F29}
\end{figure}

\begin{figure}
\centering
    \includegraphics[scale=0.4]{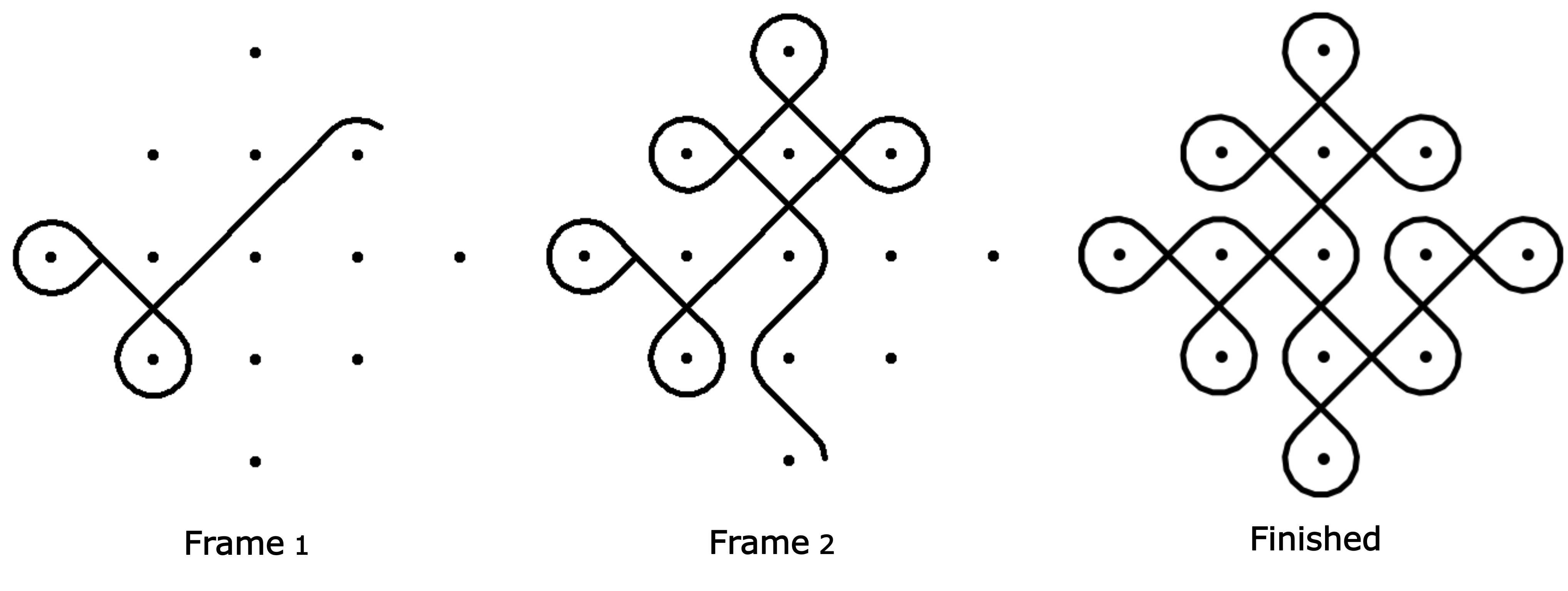}
    \caption{A $1-5-1$ rhombic Kolam.}
\label{F30}
\end{figure}

\begin{figure}
\centering
    \includegraphics[scale=0.3]{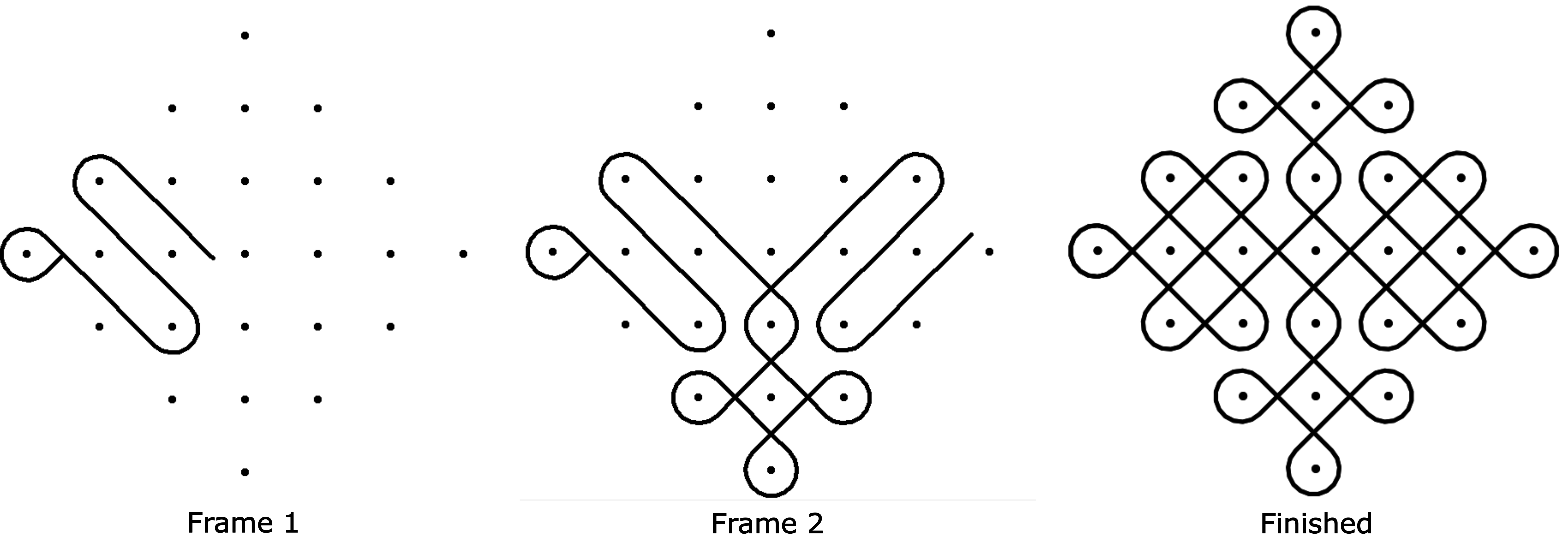}
    \caption{A $1-7-1$ rhombic Kolam.}
\label{F31}
\end{figure}

We simulated nine Navagraha ($3\times3$ square) Kolams, one $1-3-1$ Kolam, one $1-5-1$ Kolam and one $1-7-1$ Kolam using this methodology in Python using the Turtle library. The Navagraha  Kolam digital sequences are given in Table \ref{T2} and the rhombic Kolam sequences are given in Table \ref{T3}. The Kolam drawing outputs are presented in Figures \ref{F20}-\ref{F31}.\par
Each figure shows two images taken while simulating the Kolam and an image of the Kolam design after its final completion. \par

Real-life Kolams can be physically drawn by starting at any point and in any direction. But in our proposed methodology, we have used a single starting position and a single starting direction ($135$°) for Kolam encoding. The proposed methodology is successful in the simulation of drawing the Kolams by using angles at the lattice points of the lattice structure imposed on a Kolam dot pattern. More known Kolams can be encoded and added to the library. 


\section{Conclusion}
Kolams are fascinating geometrical designs with a very rich cultural and traditional relevance. In this work, we have focused on single-loop dot Kolams which start and end at the same point (also known as \textit{Brahmamudi}). We have mimicked the human act of Kolam drawing, which entails decision-making at each lattice point of the lattice structure imposed on a Kolam dot grid. The decision-rule for proceeding is based on anticipating the next three moves required.  We propose a novel encoding scheme by mapping the angular movements of the Kolam at the lattice points into a sequence containing exactly four distinct symbols. The decision-rules are mapped onto the distinct permutations of lengths $2$, $3$ and $4$ containing 4 distinct symbols. This methodology is intuitive and helps us to easily place the known \textit{Brahmamudi} Kolams in the digital space and to simulate Kolam drawing for educational and practical purposes. We propose to investigate modern digital applications of Kolam in the area of data encoding and cryptography in our future work.

\section*{Acknowledgments}
The work presented in this paper was supported by a grant of the Ministry of Education, Govt. of India, under the Indian Knowledge Systems (IKS) Competitive Research Program, 2021.


\begin{thebibliography}{10}

\bibitem{Ascher2002}
Marcia Ascher.
\newblock The kolam tradition.
\newblock {\em American Scientist}, 90(1):56--63, 2002.

\bibitem{Akhilesh}
Akhilesh Kumar and Shailaja~D Sharma.
\newblock Survey of computational methods in kolam.
\newblock Paper presented at HOMI Young Scholars' Conference, IIT Gandhinagar,
  February 2021.

\bibitem{SIROMONEY197463}
Gift Siromoney, Rani Siromoney, and Kamala Krithivasan.
\newblock Array grammars and kolam.
\newblock {\em Computer Graphics and Image Processing}, 3(1):63--82, 1974.

\bibitem{SIROMONEY1973447}
Gift Siromoney, Rani Siromoney, and Kamala Krithivasan.
\newblock Picture languages with array rewriting rules.
\newblock {\em Information and Control}, 22(5):447--470, 1973.

\bibitem{Yanagisawa2007}
Kiwamu Yanagisawa and Shojiro Nagata.
\newblock Fundamental study on design system of kolam pattern.
\newblock {\em Bulletin of the Society for Science on Form}, 21:133--134, 2007.

\bibitem{Nagata2023}
Shojiro Nagata.
\newblock Traditional kolam patterns: Formation, symmetry and fractal nature.
\newblock In {\em The Computation Meme: Computational Thinking in the Indic
  Tradition, K Gopinath and Shailaja D Sharma (Eds)}. IISc Press, Bangalore,
  (forthcoming) 2023.

\bibitem{LINDENMAYER1968280}
Aristid Lindenmayer.
\newblock Mathematical models for cellular interactions in development i.
  filaments with one-sided inputs.
\newblock {\em Journal of Theoretical Biology}, 18(3):280--299, 1968.

\bibitem{Nagata}
Shojiro Nagata.
\newblock How many loops kolam loop pattern consists of.
\newblock {\em Forma}, January 2015.

\bibitem{John}
John Layard.
\newblock Labyrinth ritual in south india: Threshold and tattoo designs.
\newblock {\em Folklore}, 48(2):115--182, 1937.

\end{thebibliography}

\bibliographystyle{unsrt}

\end{document}